\documentclass[conference]{IEEEtran}
\IEEEoverridecommandlockouts
\usepackage{cite}
\usepackage{amsmath,amssymb,amsfonts}
\usepackage{algorithmic}
\usepackage{graphicx}
\usepackage{textcomp}
\usepackage{xcolor}
\usepackage{microtype}
\usepackage{hyperref}
\usepackage{subcaption}
\usepackage{mathtools}
\usepackage{stmaryrd}
\usepackage{multirow}
\usepackage{multicol}
\usepackage{float}
\usepackage[nofillcomment,linesnumbered,ruled,vlined,noend]{algorithm2e} 

\usepackage{sidecap}
\sidecaptionvpos{figure}{c}

\usepackage[english]{babel}
\addto\extrasenglish{
    
}
\addto\extrasenglish{
    
}

\usepackage{listings}
\usepackage{amsfonts}
\definecolor{keywordcolor}{rgb}{0.5,0,0.5}
\definecolor{textgray}{gray}{0.4}
\definecolor{mygray}{rgb}{0.5,0.5,0.5}
\newcommand{\mT}{\mathcal{T}}
\newcommand{\mM}{\mathcal{M}}

\newcommand{\mU}{\mathbb{U}}

\newcommand{\mapscript}[3]{%
    \operatorname{%
            {\vphantom{#2}}_{#1}%
        \kern-\scriptspace
        \mathnormal{#2}_{#3}%
    }%
}
\newcommand{\tdistname}[3]{#1\mapscript{#2}{\mapsto}{#3}\mM}
\newcommand{\tdist}[2]{\tdistname{\mT}{#1}{#2}}

\newcommand{\tdistfused}[4]{#1\mapscript{#2}{\xmapsto{#3}}{#4}\mM}

\newcommand{\code}[1]{\lstinline[basicstyle=\ttfamily\footnotesize,keywords={}]{#1}}

\newcommand{\dense}{\code{Dense}}
\newcommand{\compressed}{\code{Compressed}}
\newcommand{\crd}{\code{crd}}
\newcommand{\pos}{\code{pos}}

\def\BibTeX{{\rm B\kern-.05em{\sc i\kern-.025em b}\kern-.08em
    T\kern-.1667em\lower.7ex\hbox{E}\kern-.125emX}}

\usepackage{tikz}

\usetikzlibrary{positioning,decorations.pathreplacing,fit}
\newcommand{\tikzmark}[2][]{%
  \tikz[remember picture,overlay,baseline=-.5ex] \node[#1] (#2) {};%
}
\newcommand*{\drawBrace}[4][0pt]{%
    \node[draw=none, fit={(#2) (#3)}, inner sep=0pt] (rectg) {};%
    \draw [decoration={brace,amplitude=0.3em},decorate,thick,mygray]%
      ([xshift=#1]rectg.north east) --%
      coordinate[right=1em, midway] (#2#3)
      ([xshift=#1]rectg.south east);%
    \node[right=-0.75em of #2#3] (#2#3-comment) {\textcolor{mygray}{#4}};
}%

\begin{document}

\title{\name{}: Compiling Distributed Sparse Tensor Computations}


\author{\IEEEauthorblockN{Rohan Yadav}
\IEEEauthorblockA{
\textit{Stanford University}\\
Stanford, CA\\
rohany@cs.stanford.edu}
\and
\IEEEauthorblockN{Alex Aiken}
\IEEEauthorblockA{
\textit{Stanford University}\\
Stanford, CA\\
aiken@cs.stanford.edu}
\and
\IEEEauthorblockN{Fredrik Kjolstad}
\IEEEauthorblockA{
\textit{Stanford University}\\
Stanford, CA\\
kjolstad@cs.stanford.edu}
}

\definecolor{todocolor}{rgb}{0.8,0,0}
\definecolor{editcolor}{rgb}{0,0,0.8}

\newcommand{\TODO}[1]{{\color{todocolor}#1}}
\newcommand{\name}{SpDISTAL}
\newcommand{\ignore}[1]{}
\newcommand{\EDIT}[1]{#1}

\maketitle


\begin{abstract}
We introduce \name{}, a compiler for sparse tensor algebra
that targets distributed systems.
\name{} combines separate descriptions of tensor algebra expressions, sparse data structures, 
data distribution, and computation distribution.
Thus, it enables distributed execution of sparse tensor algebra expressions
with a wide variety of sparse data structures and data distributions.
\name{} is implemented as a C++ library that targets a distributed task-based runtime system
and can generate code for nodes with both multi-core CPUs and multiple GPUs.
%
\name{} generates distributed code that
achieves performance competitive with hand-written distributed
functions for specific sparse tensor algebra expressions and that
outperforms general interpretation-based systems by one to two orders of magnitude.
\end{abstract}

\section{Introduction}

Sparse tensor algebra is ubiquitous and has applications 
across many fields, including scientific computing, machine learning, and data analytics~\cite{lottery-ticket, dnn-sparsity, anandkumar-tensor-decomp, bader-tensor-app, kolecki-tensors}.
These application domains operate on ever-increasing amounts of data, and can 
benefit from the growing compute and memory offered by modern distributed machines.
However, efficiently utilizing distributed machines for sparse computations remains difficult.

We present \name{}, a system that compiles sparse tensor algebra 
to distributed machines.
\name{} allows for independent descriptions of how data and computation
should be mapped onto the memories and processors of a target machine.
\autoref{fig:lang-sample} shows C++ code implementing a distributed SpMV using \name{}.
Lines \ref{fig:lang-sample:line:format-language-begin}-\ref{fig:lang-sample:line:format-language-end} describe the sparse 
format and data distributions of tensors through a \emph{format language},
and lines \ref{fig:lang-sample:line:sched-begin}-\ref{fig:lang-sample:line:sched-end} describe a 
row-based distribution strategy through a \emph{scheduling language}.
These separate languages for data and computation allow for independently changing the
data format or distribution and the algorithm to distribute the computation.

Efficient kernel implementations for sparse tensor algebra operations can be complex,
even on a single thread. 
Distributing these computations makes it even harder to ensure correctness and performance.
There are two modern approaches to tackling this complexity: a library of kernels and interpretation.
Examples of libraries of hand-written kernels include PETSc~\cite{petsc-web-page, petsc-user-ref, petsc-efficient} and 
Trilinos~\cite{trilinos-website,tpetra-website}.
An example of an interpretation-based system is
the Cyclops Tensor Framework (CTF)~\cite{ctf-main, ctf-sparse}.

Library approaches, such as PETSc and Trilinos, implement a predefined 
set of operations, each with a fixed data format and distribution strategy.
These systems provide bare-metal performance but are inflexible in the 
face of three sources of variability.
First, the fixed set of implementations in a library means that
the implementation of some the countably infinite set of tensor 
algebra expressions using these 
building blocks will be suboptimal, and in practice there are
important computations that incur such a performance penalty.
%
%
Second, when users have a data distribution or data format not
directly supported by the system, they must reshape their data to fit
the interface, incurring significant cost.
Finally, the library approach is difficult to adapt to new hardware,
as each kernel must be rewritten and re-tuned for each new platform.

Interpreted approaches, such as CTF, execute a tensor algebra expression using
a series of distributed matrix multiplication and transposition operations.
This approach can implement all tensor algebra expressions,
but cannot achieve optimal performance for all expressions.
Our experiments show that the interpreted approach can be one to two
orders of magnitude slower than hand-tuned implementations due to
unnecessary data reorganization and communication.
%
%
Through compilation, \name{} achieves the benefits of both 
approaches, supporting the full generality of sparse tensor algebra but also
specializing implementations to the desired computation and data layouts.

The recent work of DISTAL~\cite{distal} introduced separate scheduling and 
data distribution languages for distributed dense tensor algebra compilation, 
making it capable of expressing a large variety of 
dense tensor algebra algorithms.
However, DISTAL has no notion of sparsity in its language or implementation.

We address adding sparsity into DISTAL's programming
model in a way that is separated from but also composes well with the
scheduling and data distribution languages, allowing \name{} to express a 
wide range of distributed sparse tensor algebra algorithms.
%
%
We show how to add sparsity to DISTAL's 
scheduling and data distribution languages 
in a way that is expressive and composable.
We then show how to leverage \emph{dependent partitioning}~\cite{deppart} 
to translate tensor algebra expressions and data distribution declarations with sparsity
specifications into assignments of specific sub-tensors to a distributed machine.

\EDIT{
Figure~\ref{fig:lang-sample} demonstrates how sparsity is integrated into DISTAL's
programming model in an encapsulated manner.
Line~\ref{fig:lang-sample:line:spdistal-change} declares that the matrix in a matrix-vector
multiplication is sparse. 
Apart from the sparsity annotation, the remainder of Figure~\ref{fig:lang-sample} is a valid DISTAL
program that implements a row-wise distributed matrix-vector multiplication.
\name{} allows for independent description of the sparsity structure of tensors to yield distributed
sparse tensor computations.
}

We implement \name{} by extending the DISTAL~\cite{distal} and TACO~\cite{taco} compilers
and targeting the Legion~\cite{legion} distributed runtime system.
%
%
\name{} implements sparsity in the model through a combination of compilation
techniques and runtime analysis, avoiding the need to perform complex control and 
data flow analyses over imperative code.

%
%
%

The specific contributions of this work are:
\begin{enumerate}
    \item A programming model that separates data distribution and computation distribution for sparse tensors,
    \item a data structure specification language separated from data distribution specifications, and
    \item compilation techniques to support distribution of sparse tensor algebra computations.
\end{enumerate}

We evaluate \name{} by comparing against the state-of-the-art distributed sparse 
linear and tensor algebra libraries PETSc, Trilinos and CTF on matrices
and tensors from the SuiteSparse~\cite{suitesparse}, FROSTT~\cite{frosttdataset} and
Freebase~\cite{freebase} datasets.
We find that \name{} is competitive with PETSc and Trilinos, and can outperform 
CTF by over an order of magnitude.

\begin{figure}
  \centering
  \begin{lstlisting}[frame=none,language=C++,escapechar=|,commentstyle=\color{gray},numbers=left,keywords={Format,Tensor,IndexVar,int,Machine,Grid,Param,Distribution,double,MappingVar,divide,communicate,substitute,reorder,DistributedGPU,CuBLAS,GeMM,distribute,schedule,compile,Dense,Compressed,split,parallelize, DistVar},label={lst:lang-sample},basicstyle=\ttfamily\scriptsize,xleftmargin=0.3cm]
// Declare input parameters for generated code.|\label{fig:lang-sample:line:params}|
Param pieces, n, m;
// Define the machine M as a 1D grid of processors.
Machine M(Grid(pieces));|\label{fig:lang-sample:line:machine}|

// Define the data structure and distribution for 
// each tensor. We define two dense vector formats, 
// one blocked onto M, the other replicated onto all
// processors in M. Finally, we define a CSR matrix
// format, distributed row-wise. The format notation 
// is discussed in Subsection II-B.
DistVar x, y;|\label{fig:lang-sample:line:format-language-begin}|
Format BlockedDense({Dense}, Distribution({x}, M, {x}));
Format ReplDense({Dense}, Distribution({x}, M, {y}));
Format BlockedCSR({Dense, Compressed}, 
                         Distribution({x, y}, M, {x}));|\label{fig:lang-sample:line:spdistal-change}|

// Create our tensors, using the defined formats. Our
// SpMV algorithm will block a and B, and replicate c.
Tensor<double> a({n}, BlockedDense);
Tensor<double> B({n, m}, BlockedCSR);
Tensor<double> c({m}, ReplDense);|\label{fig:lang-sample:line:format-language-end}|

// Declare the computation, a matrix-vector multiply.
IndexVar i, j;
a(i) = B(i, j) * c(j);|\label{fig:lang-sample:line:tensor-index-notation}|

// Map the computation onto M via scheduling commands.
IndexVar io, ii;
a.schedule()|\label{fig:lang-sample:line:sched-begin}|
 // Block i for each node.
 .divide(i, io, ii, M.x)|\label{fig:lang-sample:line:dist-start}|
 // Distribute each block of i onto each node.
 .distribute(io)|\label{fig:lang-sample:line:dist-end}|
 // Communicate the needed sub-tensor for each chunk of i.
 .communicate({a, B, c}, io)
 // Schedule the leaf computation that runs on each node.
 // Here, we parallelize chunks of i over CPU threads.
 .parallelize(ii, CPUThread)|\label{fig:lang-sample:line:sched-end}|;
  \end{lstlisting}
  \caption{Distributed CPU SpMV kernel in \name{}.}
  \label{fig:lang-sample}
\end{figure}


\begin{figure}
    \centering
    \includegraphics[width=0.9\linewidth]{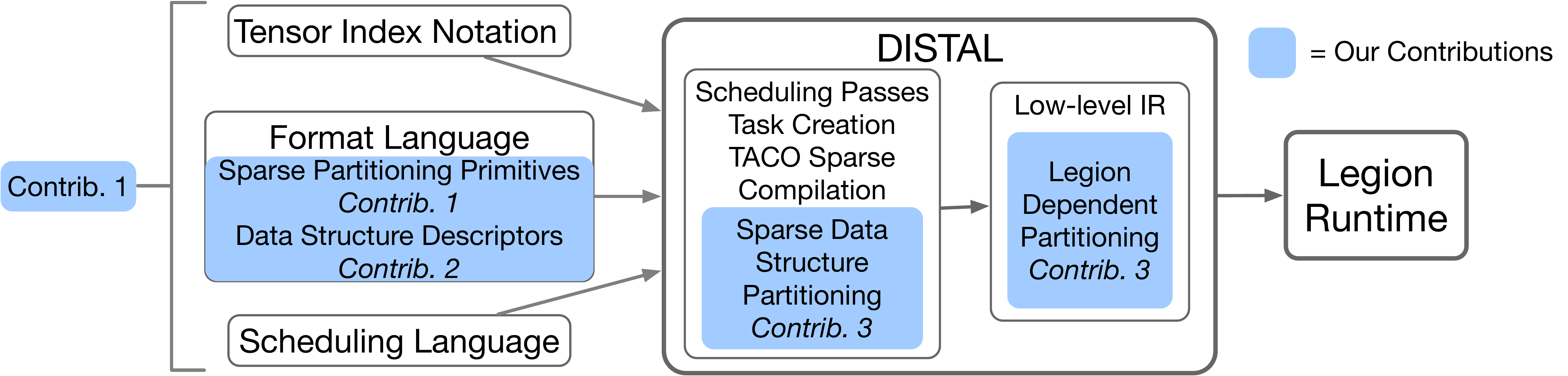}
    \caption{Overview of \name{}'s contributions.}
    \label{fig:contributions}
\end{figure}

\section{Programming Model}\label{sec:programming-model}

The first contribution of \name{} is a programming model that combines sparse data structure
specifications with separate data and computation distribution languages.
\autoref{fig:lang-sample} utilizes the three input sub-languages of \name{} to implement
a distributed SpMV: a \emph{computation language} that describes the desired kernel (line \ref{fig:lang-sample:line:tensor-index-notation}),
a \emph{format language} (lines \ref{fig:lang-sample:line:format-language-begin}-\ref{fig:lang-sample:line:format-language-end}) 
that describes how input tensors store non-zeros and are distributed onto a machine,
and a \emph{scheduling language} (lines \ref{fig:lang-sample:line:sched-begin}-\ref{fig:lang-sample:line:sched-end})
that describes how to optimize and distribute the computation.
\EDIT{
Components of these input languages have been proposed by prior works, as discussed
in the next few sections.
However, the first key contribution of \name{} is the novel combination of these languages 
to support distributed sparse tensor algebra.
}

\subsection{Computation Language}

Computation is described in \name{} using {\em tensor index notation} (TIN).
We adopt the concrete syntax of TACO~\cite{taco} and DISTAL~\cite{distal} for TIN.
%
TIN consists of \emph{accesses} that index tensor dimensions with lists of \emph{index variables}.
TIN statements are assignments into a left-hand side access, while the right
hand side is an expression constructed from multiplication and additions between any number
of accesses.
For example, the tensor-times-vector operation is expressed in TIN
as $A(i, j) = B(i, j, k) \cdot c(k)$, declaring that each element $A(i, j)$ is the inner product
between the final dimension of $B$ and the vector $c$.
Intuitively, each distinct index variable corresponds to a loop, and variables contained only 
in the right-hand side of the assignment represent sum-reductions over their domain.
%

\subsection{Format Language}\label{sec:prog-model:format-language}
\begin{figure}
    \centering
    \includegraphics[width=0.95\linewidth]{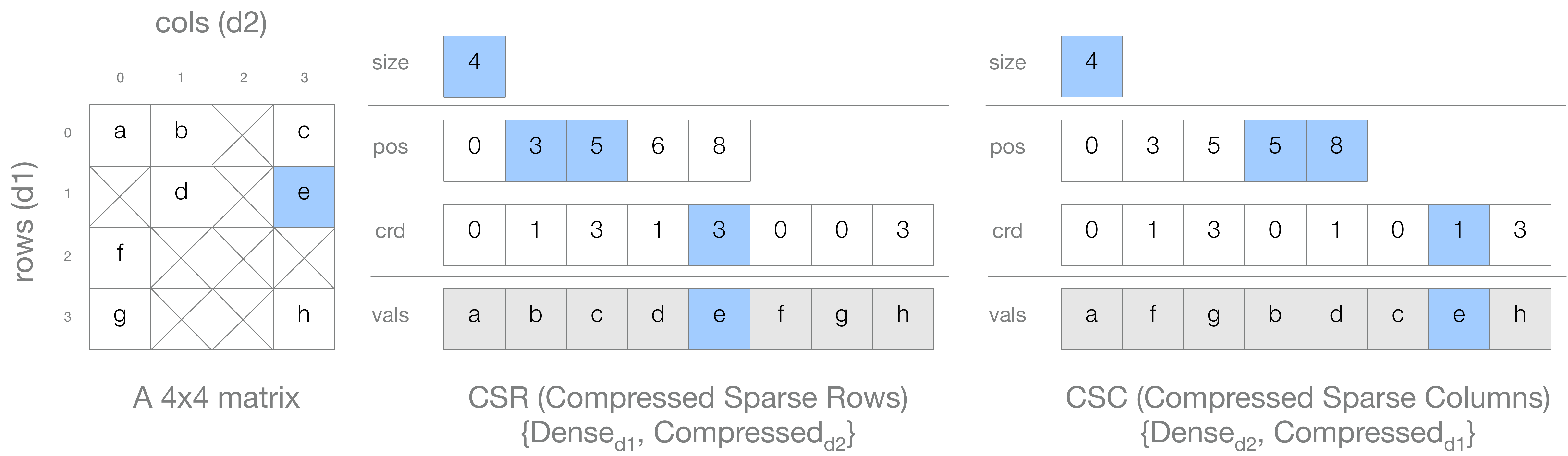}
    \caption{Different encodings of a sparse matrix \EDIT{in TACO's format language}. 
    Blue squares indicate how the coordinate (1,3) is represented
    by each encoding.}
    \label{fig:taco-formats}
\end{figure}

\emph{Sparse Data Structures.}
\name{} adopts the format language of TACO~\cite{taco}, letting users 
control tensor data structures and thus how zero entries are compressed.
TACO's format language allows users to specify the format used to store each 
dimension of a sparse tensor.
Two formats considered in TACO are the \dense{} and \compressed{} formats.
A k-dimensional tensor is stored using any combination of k instances of these formats.
A \dense{} format represents a standard array containing all coordinates 
of the dimension.
A \compressed{} format encodes only the non-zero
coordinates of the dimension using two arrays: a \crd{} array
that stores the non-zero coordinate values and a \pos{} array
that stores the range of coordinates associated with each entry in the
previous dimension.
\autoref{fig:taco-formats} shows how \EDIT{TACO's} per-dimension approach allows for
the expression of a variety of different common formats.
For example, the CSR matrix encoding (shown in the center) is constructed by using a
\dense{} encoding of the first dimension and a \compressed{}
encoding of the second dimension.
The CSC matrix encoding (shown on the right) instead uses a \dense{} encoding of the second
dimension and a \compressed{} encoding of the first dimension, and orders
the dimensions in reverse.
%
%
%

\begin{figure}
    \begin{subfigure}[b]{0.49\linewidth}
        \centering
        \includegraphics[width=0.4\textwidth]{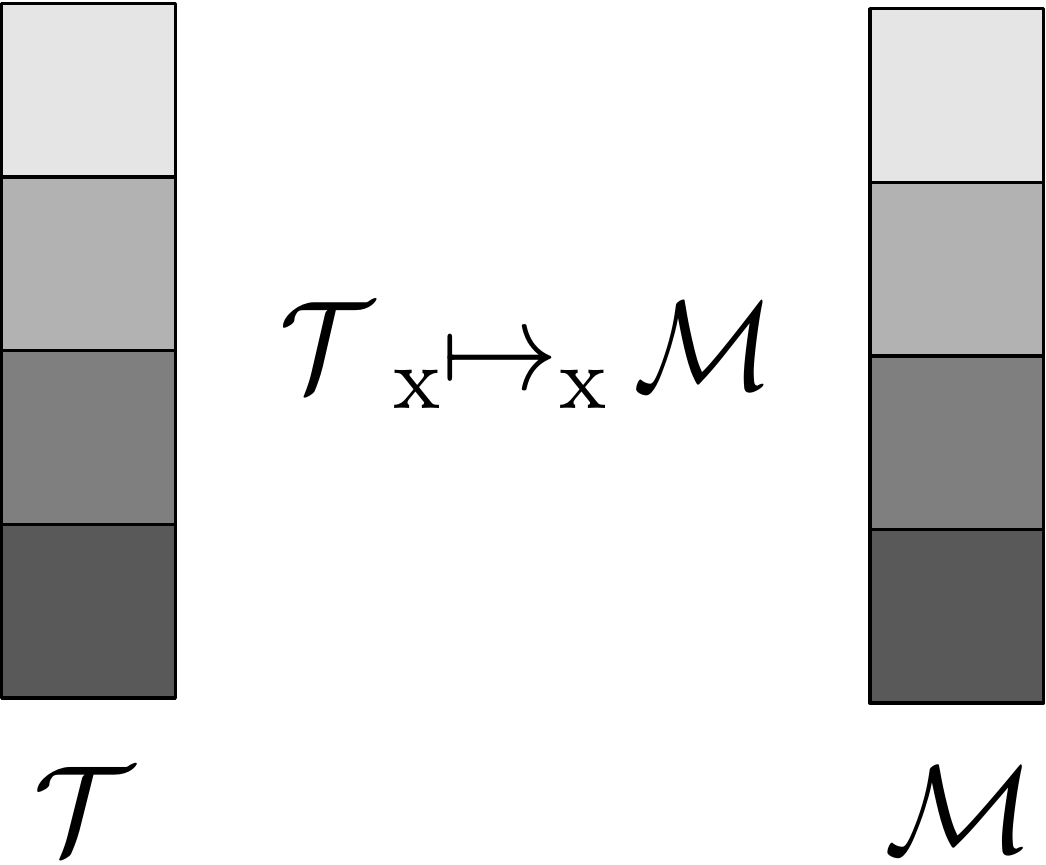}
        \captionof{figure}{Blocked vector distribution.}
        \label{fig:ti_mt_mi}
    \end{subfigure}%
    \begin{subfigure}[b]{0.49\linewidth}
        \centering
        \includegraphics[width=0.6\textwidth]{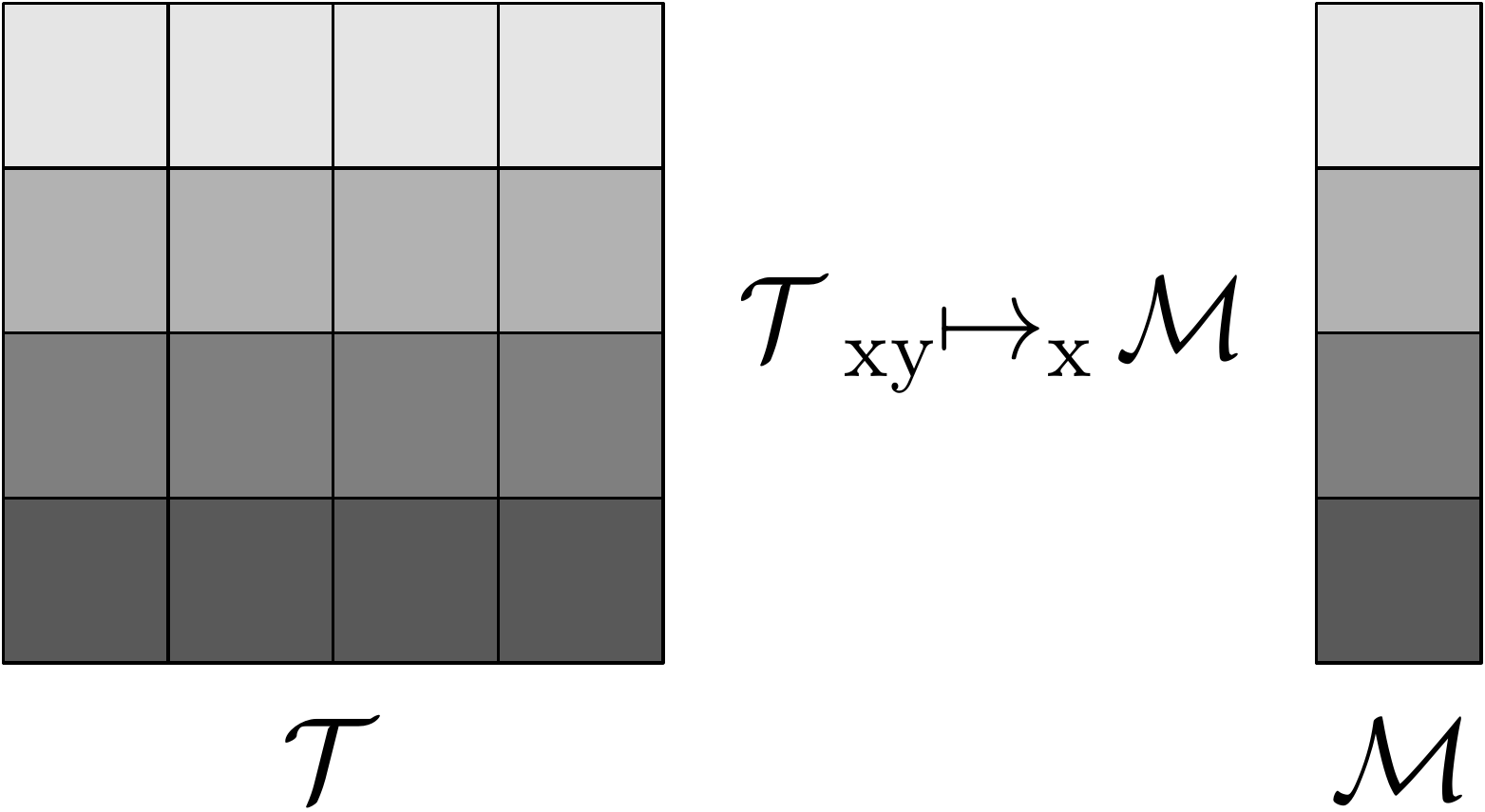}
        \captionof{figure}{Row-wise matrix distribution.}
        \label{fig:tij_mt_mi}
    \end{subfigure}
    
    \par\bigskip
    
    \begin{subfigure}[b]{0.49\linewidth}
        \centering
        \includegraphics[width=0.8\textwidth]{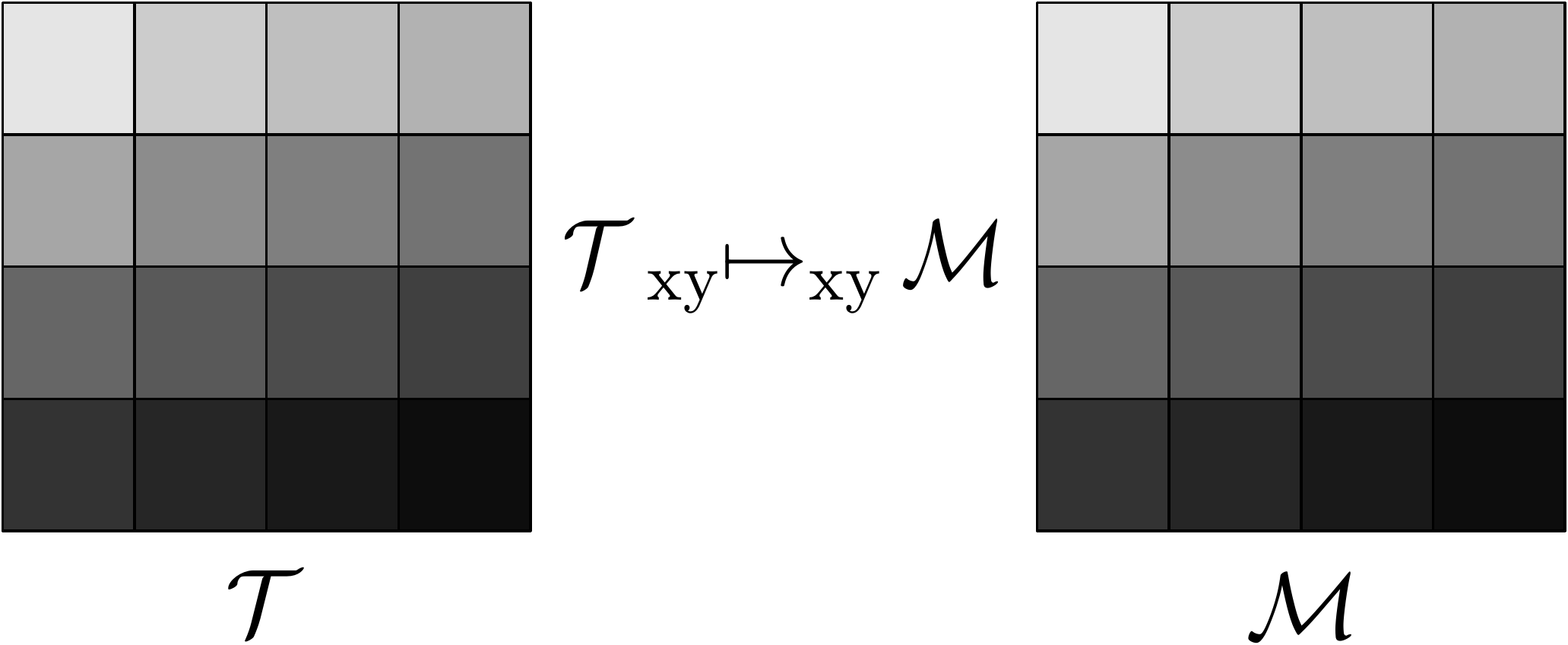}
        \captionof{figure}{Tiled matrix distribution.}
        \label{fig:tij_mt_mij}
    \end{subfigure}%
    \begin{subfigure}[b]{0.49\linewidth}
        \centering
        \includegraphics[width=0.8\textwidth]{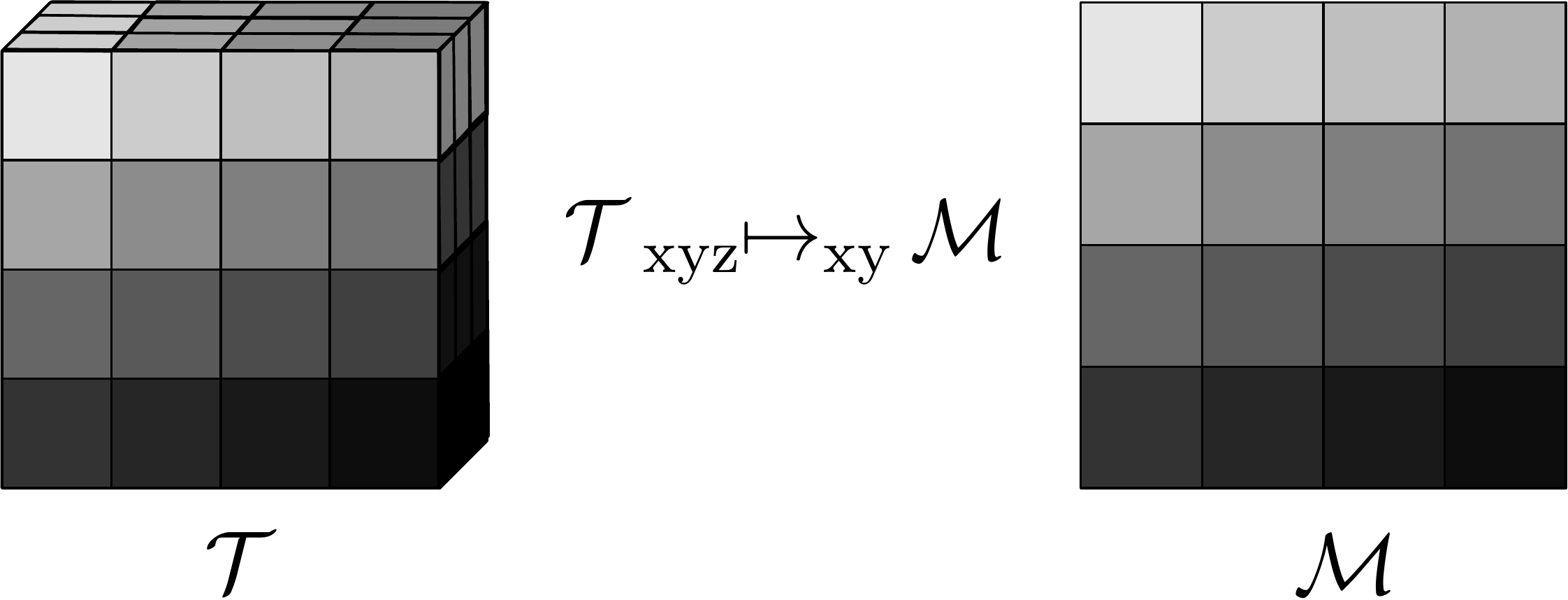}
        \captionof{figure}{Map 3-tensor onto proc. grid.}
        \label{fig:tij_mt_mij}
    \end{subfigure}
    \caption{Examples of tensor distribution notation statements.\protect\footnotemark}
    \label{fig:example-distal-data-distributions}
\end{figure}

\emph{Data Distribution.}
\name{} extends the \emph{tensor distribution notation} (TDN) language
developed in DISTAL~\cite{distal} with new constructs for describing distributions
of sparse tensors.
TDN lets users specify how each dimension of a tensor is partitioned onto 
different dimensions of an abstract machine represented by an $n$-dimensional grid.
A TDN statement assigns names to each dimension of a tensor
and a machine, and tensor dimensions that share a name with a machine dimension
are partitioned by the corresponding machine dimension.
For example, the TDN statement $\tdist{xy}{x}$ maps a matrix $\mT$ row-wise onto
a one dimensional machine $\mM$, declaring that the first dimension of $\mT$ 
is partitioned by the first dimension of $\mM$.
We include four examples of TDN statements for dense tensors in \autoref{fig:example-distal-data-distributions}.

\footnotetext{Figures used with author permission from Yadav et al.~\cite{distal}}

\begin{figure}
    \begin{subfigure}[c]{0.48\linewidth}
        \begin{subfigure}[c]{\linewidth}
            \centering
            \includegraphics[width=0.7\linewidth]{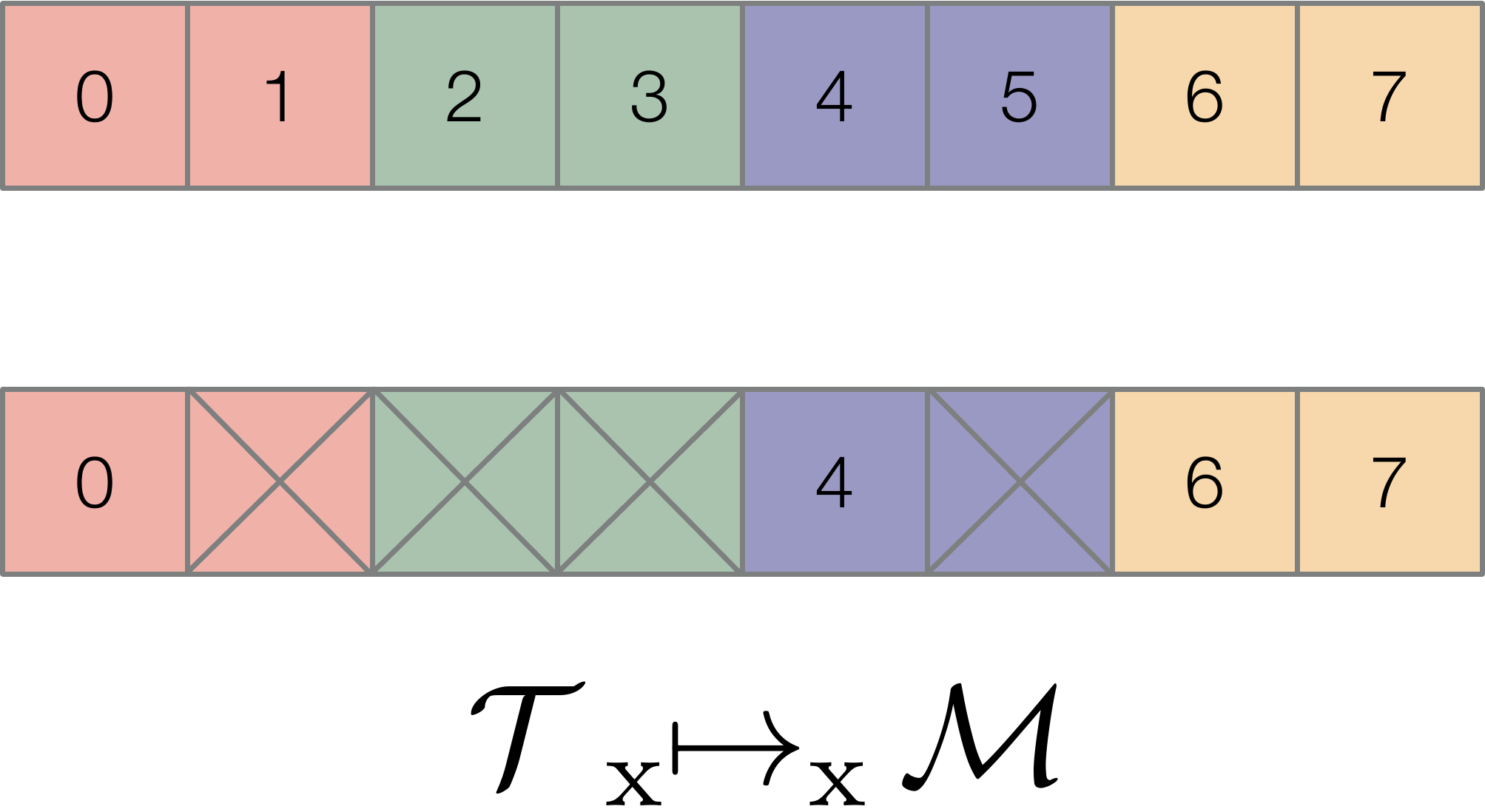}
            \caption{Universe partitions of dense and sparse vectors.}
            \label{fig:vec-univ-part}
        \end{subfigure}
        \par\bigskip
        \begin{subfigure}[c]{\linewidth}
            \centering
            \includegraphics[width=0.7\linewidth]{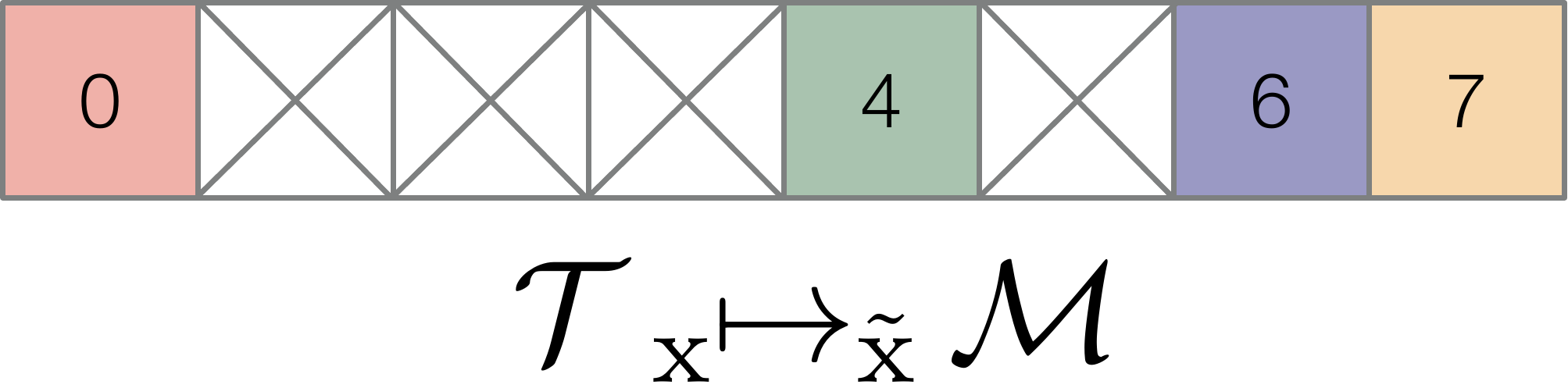}
            \caption{Non-zero partition of a sparse vector.}
            \label{fig:vec-coord-part}
        \end{subfigure}
    \end{subfigure}\hfill
    \begin{subfigure}[c]{0.48\linewidth}
        \centering
        \includegraphics[width=0.5\linewidth]{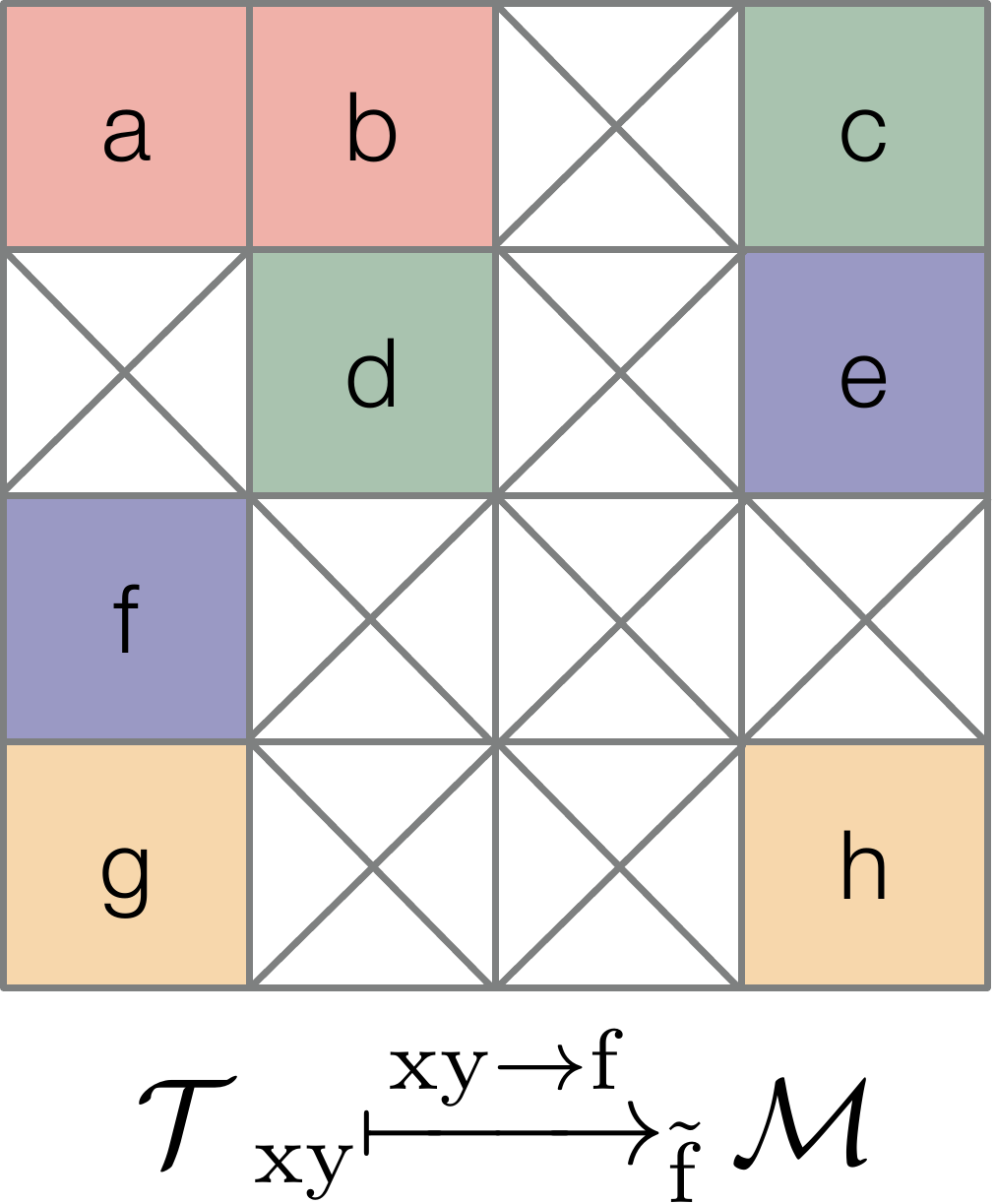}
        \caption{Fused non-zero partition of sparse matrix.}
        \label{fig:fused-coord-part-mat}
    \end{subfigure}
    \caption{Sparse tensors distributed with TDN. Each color indicates a group of coordinates mapped to the same processor.}
    \label{fig:sparse-parts}
\end{figure}

In \name{}, we extend TDN with \emph{universe} and \emph{non-zero} partitions, 
and \emph{coordinate fusion}.
TDN's default partition is a \emph{universe partition}: when a tensor dimension $d$ 
is partitioned by a machine dimension, the range of coordinates of $d$ 
(i.e., the universe\footnote{We refer to the set of all coordinates of a tensor 
dimension as the universe, from which a sparse tensor may store only  a subset.} $\mU$) 
is partitioned equally among processors in that machine dimension.
As seen in \autoref{fig:vec-univ-part}, universe partitions can be 
applied to sparse tensor dimensions, partitioning the non-zero coordinates 
according to the equal partition of $\mU$.

Universe partitions of sparse tensor dimensions may result in imbalance, 
as the non-zero coordinates may not align with the universe partitions.
We therefore introduce \emph{non-zero partitions},
which declare that the non-zero coordinates of $d$ should be partitioned evenly
using the tilde operator $\tilde{d}$ on a machine dimension.
For example, the TDN statement $\tdist{x}{\tilde{x}}$ declares
that the non-zeros of a sparse vector should be distributed equally onto $\mM$ 
and is visualized in \autoref{fig:vec-coord-part}.

In isolation, non-zero partitions are not sufficient to evenly distribute
higher order tensors.
For example, a matrix with no empty rows but a varying number of non-zeros per row 
has the same distribution under a universe and non-zero
partition of the first dimension.
We therefore introduce \emph{coordinate fusion}, allowing for
operations such as evenly distributing the non-zero coordinates of a sparse matrix.
Coordinate fusion collapses multiple dimensions into a single logical dimension,
which can be the target of a non-zero partition.
The syntax $\tdistfused{\mT}{xy}{\text{xy} \rightarrow \text{f}}{\tilde{f}}$ 
utilizes coordinate fusion to equally distribute the non-zeros of a matrix 
by flattening $x$ and $y$ into a new coordinate $f$
and then performing a non-zero partition of $f$, visualized in \autoref{fig:fused-coord-part-mat}.
The combination of fusion and non-zero partitioning allows for expression of a variety
of non-trivial data partitioning strategies.
For example, the three distributions 
$\tdist{xyz}{\tilde{x}}$,
$\tdistfused{\mT}{xyz}{\text{xy} \rightarrow \text{f}}{\tilde{f}}$, and
$\tdistfused{\mT}{xyz}{\text{xyz} \rightarrow \text{f}}{\tilde{f}}$
map a 3-tensor onto a machine by equally distributing the non-zero slices,
the non-zero tubes and non-zero values respectively.
Fusion and non-zero partitions do not subsume universe
partitions---\autoref{sec:programming-model:spmv} describes
tradeoffs between the gained load balance and additional communication.

\subsection{Scheduling Language}\label{sec:scheduling-language}

Like many domain specific languages~\cite{taco_scheduling, halide, taco_workspaces, tiramisu, TVM, tensor-comprehensions, graphit}, \name{} separates the computation description from
how the computation should be executed through a \emph{scheduling language}
that describes optimizing transformations.
Common transformations introduced by prior work that \name{} uses include
\code{parallelize} (parallelize loop iterations), \code{precompute} (hoist 
computation out of a loop), \code{split/divide} 
(break a loop into two nested loops),
\code{fuse} (collapse two loops into a single loop), and 
\code{reorder} (change the order of loops).

\EDIT{
Though these common loop transformations have generally applied to dense loop nests,
Senanayake et al.~\cite{taco_scheduling} show how to extend TACO to support these
transformations on sparse iteration spaces on a single node.
To perform additional optimizations on sparse iteration spaces, Senanayake et al.
also introduce variants of the \code{split} and \code{divide} transformations
that enable the iterations over only non-zero values to be strip-mined
into equal pieces.
These variants compose with the \code{fuse} and \code{parallelize}
commands to enable statically load balanced iteration over certain sparse loops.
For example, Senanayake et al. show how to implement an SpMV that is load-balanced
over CPU threads by \code{fuse}-ing the $i$ and $j$ dimensions of the computation,
applying the non-zero based variant of \code{split}, and then \code{parallelize}-ing
the resulting outer loop.
We refer to Sections 2 and 3.3 of Senanayake et al.~\cite{taco_scheduling} for more details.
}
%
%

\EDIT{Finally, DISTAL~\cite{distal} introduced new scheduling commands
to target distributed machines,} namely \code{distribute} and
\code{communicate}.
\code{distribute} directs that iterations of the
target loop should execute on different processors.
The \code{communicate} command directs that the necessary
subsets of each target tensor should be fetched to a local memory at the beginning of each iteration of the target loop.
The \code{communicate} command 
is only used for optimization---users control the granularity of 
communication for performance and DISTAL infers what data to 
communicate and the source and destination of transfers.

\EDIT{
\name{}'s scheduling language combines the single-node sparse iteration space
transformations in TACO~\cite{taco_scheduling} with DISTAL's distributed scheduling transformations~\cite{distal}, 
enabling the distribution of sparse tensor programs.
This combination of scheduling transformations is novel and unique to \name{}.
}

%
%

\subsection{Putting It Together: Scheduling SpMV}\label{sec:programming-model:spmv}

To showcase the \name{} programming model, 
we discuss two algorithms for the SpMV kernel, $a(i) = B(i, j) \cdot c(j)$
that target a machine $\mM$ organized as one-dimensional grid of processors.
The first uses a row-based distribution, implemented using \name{}'s C++ API in \autoref{fig:lang-sample}, 
and the second uses a non-zero-based distribution of the computation.
We use the same sparse formats for each tensor in both distributed algorithms:
\code{a, c = \{Dense\}} and \code{B = \{Dense, Compressed\}}.
%
%

In the row-based algorithm, each processor is assigned rows of $B$,
the corresponding elements of $a$, and all of $c$.
To avoid data movement at compute time, we declare the data distributions of
each tensor as follows: $\tdistname{a}{x}{x}$, $\tdistname{B}{xy}{x}$ and $\tdistname{c}{x}{y}$.
To schedule this algorithm, we 
\code{divide} to
create a group of rows for each processor, and \code{distribute} the
groups over each processor.
We then use \code{communicate} on \code{a}, \code{B} and \code{c} at the distributed loop,
pre-fetching each sub-tensor at the start of processing each group of rows.

When the rows of $B$ have different numbers of non-zeros, a row-based 
algorithm can suffer from load imbalance.
An algorithm that evenly splits the non-zeros, on the other hand, 
enables perfect load balance at the cost of communication to 
reduce into the output.
This algorithm partitions the elements of $a$ (not necessarily evenly) and replicates $c$.
We choose a non-zero based distribution of $B$ using coordinate fusion
and a non-zero partition with $\tdistfused{B}{xy}{xy \rightarrow f}{\tilde{f}}$.
To schedule the computation we 
\code{fuse} the \code{i} and \code{j} loops, then 
\code{divide} the space of non-zero
coordinates of $B$.
Finally, we \code{distribute} and
\code{communicate} as before to 
complete the schedule.

In both algorithms, we matched the data and computation distributions to avoid
unnecessary communication.
However, this is not required---a \name{} program that utilized the row-based schedule
but the non-zero based data distribution
is valid but comes at a performance cost, as extra communication operations would be
needed to reshape the non-zero based data distribution into a row-based one.

\section{Distributed Sparse Tensor Data Structures}\label{sec:data-structures}

The first step to realizing the programming model discussed in \autoref{sec:programming-model}
is to define the data structures used to represent distributed sparse tensors.
In this section, we describe abstract distributed data types on which we lay out distributed sparse
tensors.
Then, in \autoref{sec:compilation}, we describe how to generate code that distributes
and computes on these sparse tensors.
\EDIT{
These abstract data structures are then implemented by a runtime system (Legion~\cite{legion})
through dynamic analyses.
}

\begin{figure}
    \begin{subfigure}[b]{0.5\linewidth}
        \centering
        \includegraphics[width=0.7\linewidth]{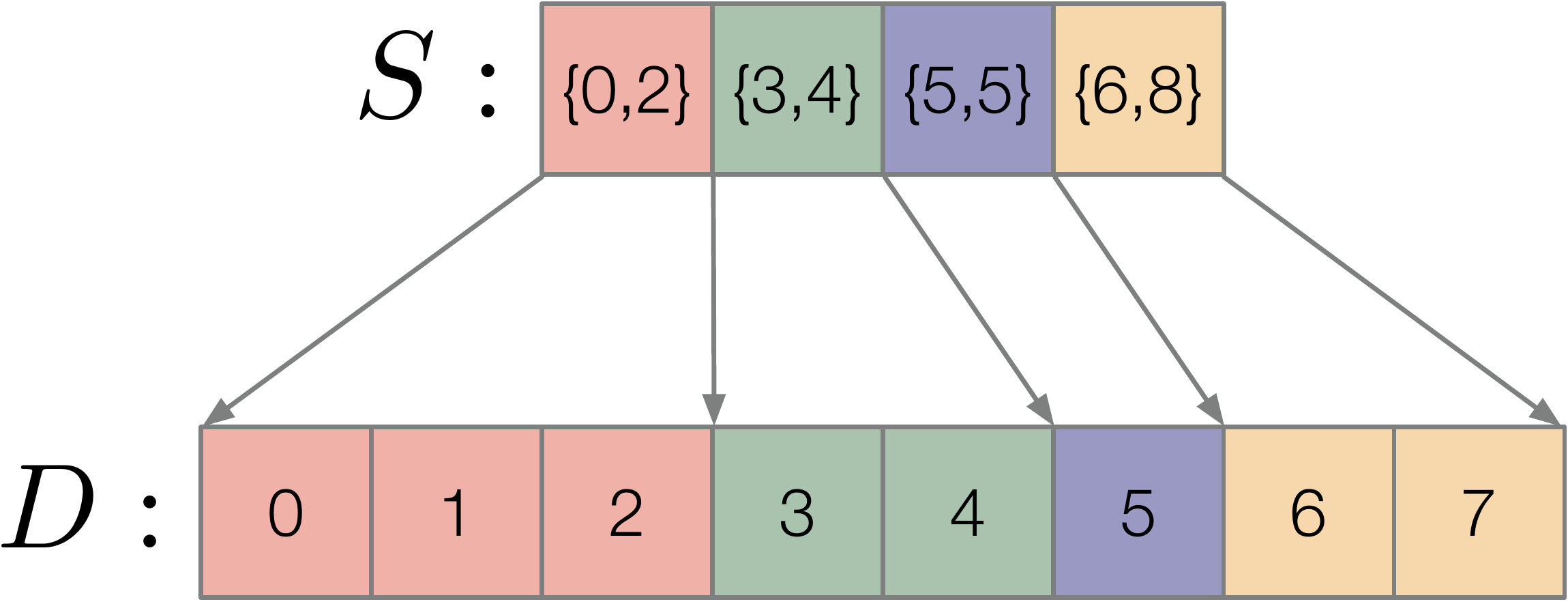}
        \caption{Image}
        \label{fig:deppart-image}
    \end{subfigure}\hfill
    \begin{subfigure}[b]{0.5\linewidth}
        \centering
        \includegraphics[width=0.7\linewidth]{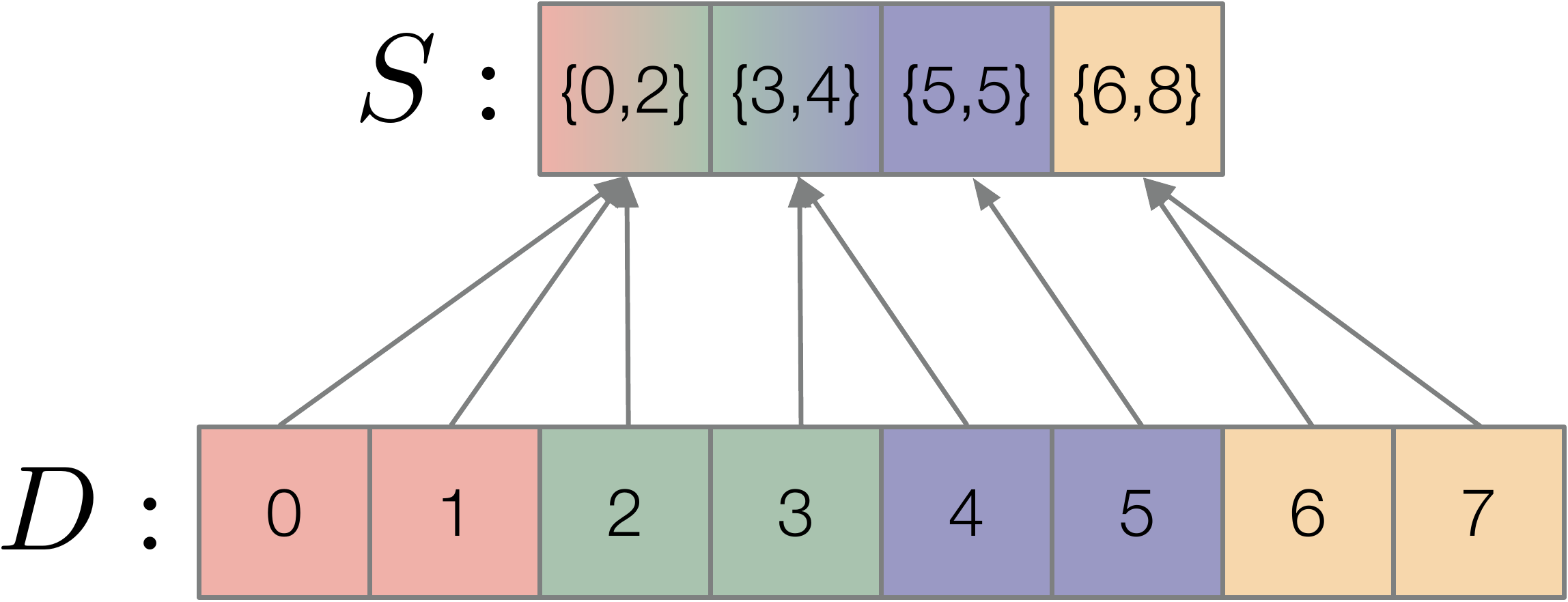}
        \caption{Preimage}
        \label{fig:deppart-preimage}
    \end{subfigure}\hfill
    \caption{Visualization of image and preimage operations.}
    \label{fig:deppart-example}
\end{figure}

\subsection{Abstract Distributed Data Structures}

To manage the complexity of distributing sparse tensors, we first reduce them
to abstract data types that more directly lend themselves to distribution.
These abstract data types come from the Legion ecosystem~\cite{legion, deppart}.

\EDIT{
An \emph{index space} is an abstract object representing a set of indices
or coordinates, where the indices can have any number of dimensions.
A \emph{region} is a multi-dimensional array of values, where 
the values can be primitive types such as integers or floats, or structured data
such as index spaces that name sets of indices in other regions.
As a multi-dimensional array, a region can be viewed as a function from indices in a 
multi-dimensional index space to a set of values.
Therefore, each region is associated with an index space that describes the valid
set of indices that can be used to access the region.

A \emph{partition} is an abstract object representing a mapping from a set
of \emph{colors} to (potentially overlapping) subsets of an index space.
We depict partitions by shading subsets of regions different colors.
Regions can be \emph{partitioned} into \emph{sub-regions} by constructing
a partition of the region's associated index space.
Each sub-region is associated
with the corresponding subset of the original partitioned index space.
Regions are distributed by partitioning them into sub-regions that are placed
onto different memories across a distributed machine.

Partitions are created either through direct coloring of subsets of indices,
or from existing partitions through a class of operations known as 
\emph{dependent partitioning}~\cite{deppart} operations.
We utilize two dependent partitions operations \emph{image} and \emph{preimage}
that are applicable to regions containing index spaces as values.
Regions with index spaces as values encode pointers to indices of other regions.
For example, in \autoref{fig:deppart-image}, the first element of the top region
is an index space corresponding to the set $\{0, 1, 2\}$, pointing to the first three
elements of the bottom region.
Intuitively, image colors all destinations of pointers with the same color as their source,
while preimage colors all sources of pointers with the same color as their destination.

We now give precise definitions of image and preimage.
Consider a source region $S$ containing index spaces that name indices in a destination
region $D$.
Given a partition $P_S$ of $S$, \texttt{image($S, P_S, D$)} is a partition
$P'$ of $D$ such that $\forall c \in P_S, \forall i \in P_S[c], S[i] \subseteq P'[c]$, depicted
in \autoref{fig:deppart-image}.
Preimage performs the inverse operation: given a partition $P_d$ of $D$, \texttt{preimage($S, P_D, D$)}
is a partition $P'$ of $S$ such that 
$\forall c \in P_D, \forall i \in P_D[c], \forall i'~\textsf{s.t.}~i \in S[i'], i' \in P'[c]$, depicted
in \autoref{fig:deppart-preimage}.
We reiterate that partitions can contain overlapping subsets of index spaces.
As seen in \autoref{fig:deppart-preimage}, the resulting partition of $S$ colors
some indices with multiple colors.
Data referenced by multiple sub-regions in a partition is shared across the different 
memories that sub-regions are placed in, and the runtime system manages coherence between
the different copies.
}

\begin{figure}
    \centering
    \includegraphics[width=\linewidth]{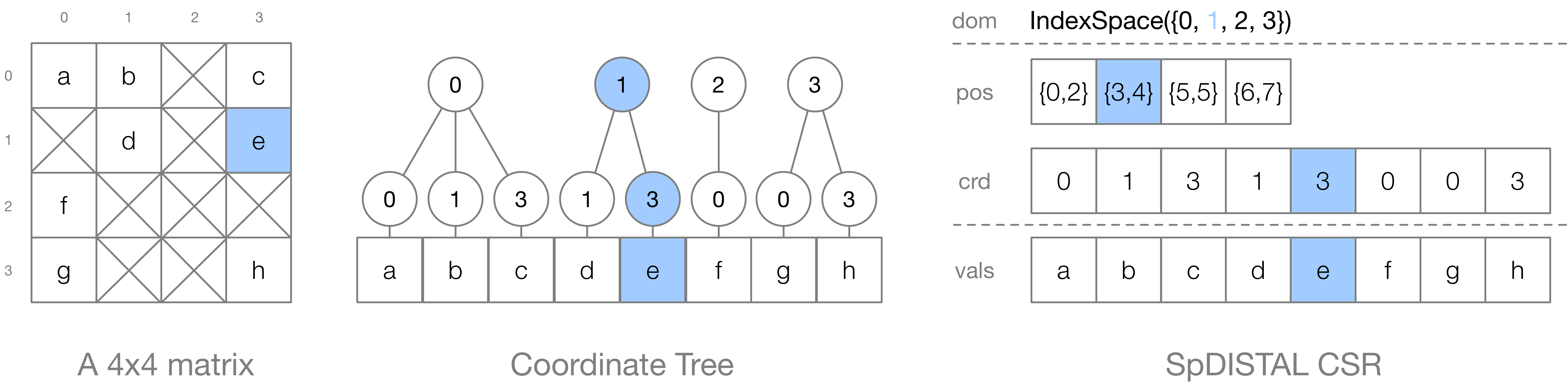}
    \caption{Coordinate tree and \name{} CSR encodings.
    }
    \label{fig:spdistal-csr}
\end{figure}

\subsection{Distributed Sparse Tensors}

Having described abstract data structures that represent distributed arrays
and operations to create expressive partitions of these arrays, we
show how to use them to represent sparse tensors.
Our distributed sparse tensor encoding is a distributed extension of the
encoding proposed by TACO~\cite{taco} and discussed in
\autoref{sec:prog-model:format-language}.
To give intuition about our (and TACO's) sparse tensor encoding, we describe
the \emph{coordinate tree} representation of a tensor.

A tensor $\mT$'s coordinate tree is a tree with one level per dimension of $\mT$
(in addition to a root), as shown in the center of \autoref{fig:spdistal-csr}.
Each path from the root to a leaf node represents a non-zero tensor coordinate.
The levels of the coordinate tree are ordered in the way the dimensions of $\mT$ are
stored.
For example, a row-major layout of a matrix would have rows followed by columns, while a
column-major layout would have columns followed by rows.
Each level of the coordinate tree is represented with a \emph{level format}, which encodes
the non-zero coordinates of the level with different data structures.

\EDIT{
Sparse tensors are encoded by specifying how each level of the coordinate tree is stored
independently of the other levels.
Each coordinate tree level is represented by a \emph{level format} that defines how
the non-zero coordinates of the coordinate tree level are represented with
different physical data structures.
In this paper, we consider two level formats: the \dense{} level format that
stores all coordinates of a level in a dense array, and the
\compressed{} level format that encodes the non-zero coordinates of a level
by compressing away the zero coordinates.

The \dense{} level format is used to encode coordinate tree levels that contain
few zero values.
}
%
%
We encode \dense{} levels with an index space
over the range from zero to the size of the dense level, named \code{dom}.
Multiple contiguous \dense{} levels are collapsed into a single
logical multi-dimensional \dense{} level that is represented by a
multi-dimensional index space.

\compressed{} levels are encoded in TACO with an array of coordinates
(\crd{}) and an array that stores bounds on the range of coordinates
associated with each entry in the parent level (\pos{}).
The coordinates for an entry $i$ in the parent level
are stored in the positions of \crd{} 
within the range \code{[pos[i], pos[i+1])}.
\name{} stores the \pos{} and \crd{} arrays as regions to enable partitioning and distribution.
As the \pos{} region contains ranges of the \crd{} region, our
goal is to utilize image and preimage to relate partitions of \pos{}
and \crd{}.
Therefore, we store tuples in the \pos{} region that contain the lower
and upper bounds of coordinate positions, representing a set of indices 
in the \crd{} region.
The coordinates for an entry $i$ in the parent level are stored in positions
\code{[pos[i].lo, pos[i].hi]} of \crd{}.
\autoref{fig:spdistal-csr} visualizes an \name{} CSR matrix.

\section{Compiling Distributed Sparse Tensor Programs}\label{sec:compilation}

Having described the data structures that represent sparse tensors, we now show how 
to generate distributed sparse tensor algebra code.
Since we can utilize single node code generation for sparse
tensor algebra~(\hspace{1sp}\cite{taco,taco_scheduling}),
the key challenge is partitioning sparse tensors
into sub-tensors that can be processed on a single node.
%
%
Describing partitions of sparse tensors enables moving sparse tensors 
into a specified distribution or to where they are needed by a computation.
Our approach has two components: 1) a set of compiler
abstractions to reason about data partitioning, and 2)
a code generation algorithm that uses these abstractions to generate partitioning
code specialized to a computation or data distribution specification.
We first provide the intuition behind our approach, and then describe the novel
compiler abstractions and code generation process.





\begin{figure}[t]
    \begin{subfigure}[b]{0.49\linewidth}
        \centering
        \includegraphics[width=0.8\linewidth]{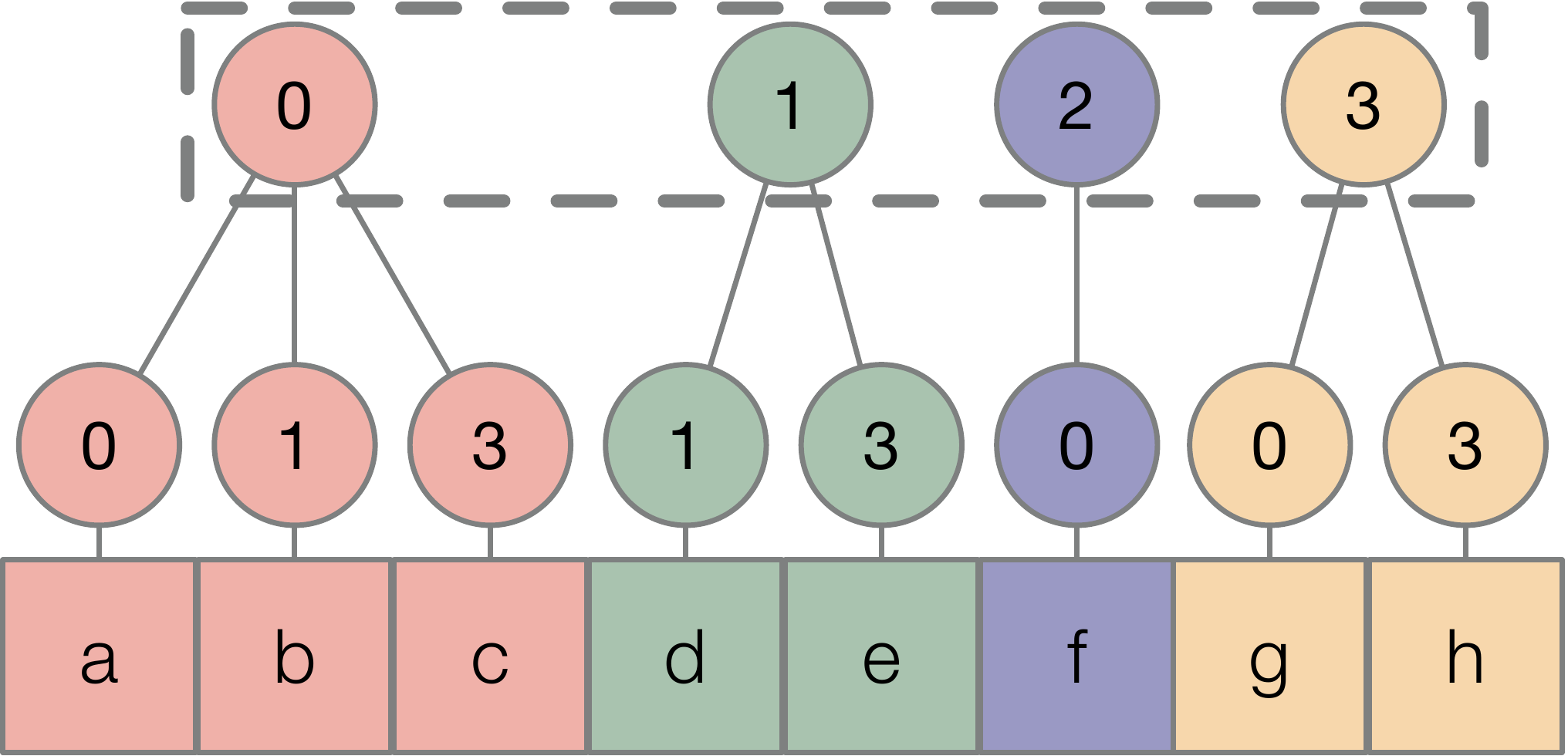}
        \caption{Universe partition of level one.}
        \label{fig:coord-tree-univ-full-part}
    \end{subfigure}\hfill
    \begin{subfigure}[b]{0.49\linewidth}
        \centering
        \includegraphics[width=0.8\linewidth]{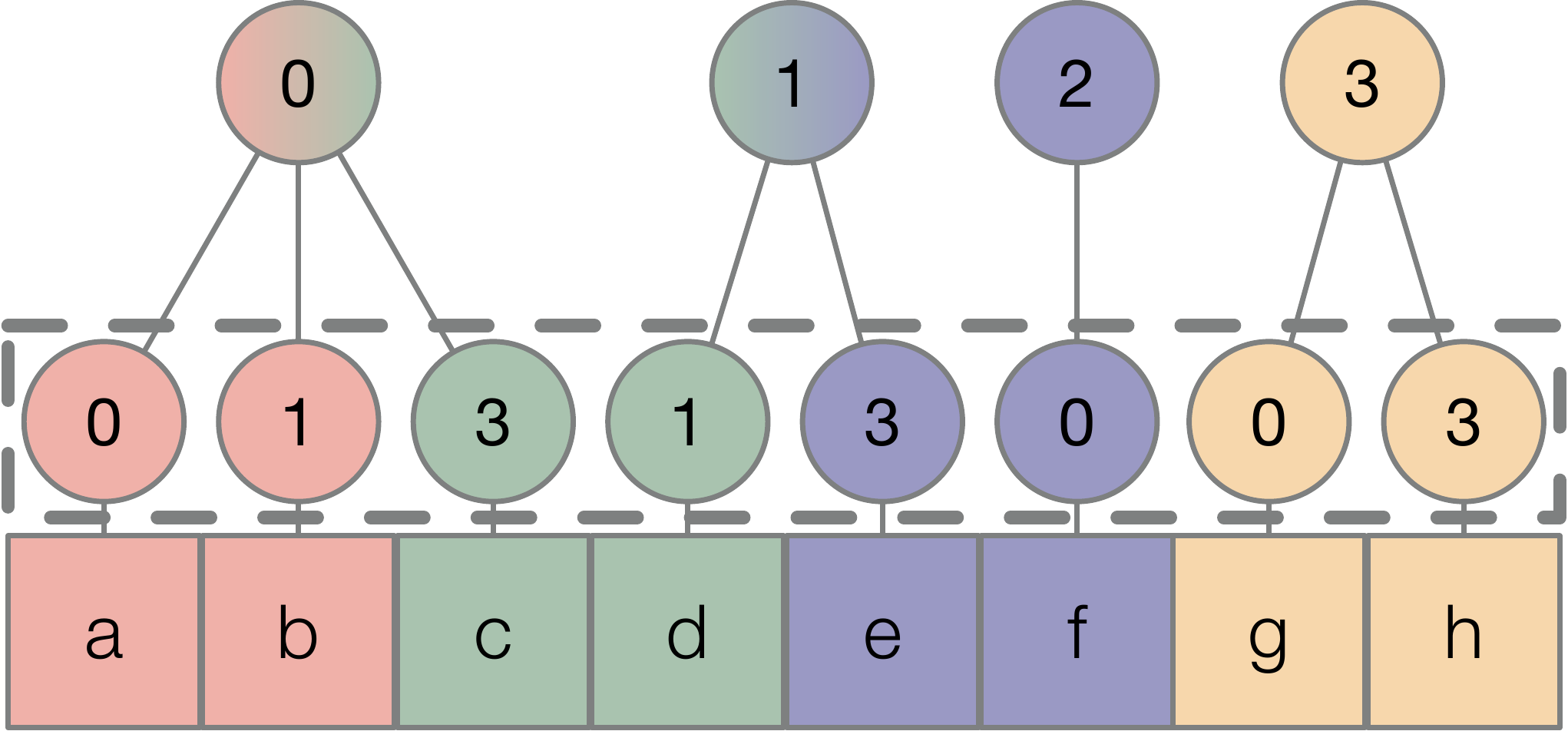}
        \caption{Non-zero partition of level two.}
        \label{fig:coord-tree-full-coord-part}
    \end{subfigure}
    \caption{Coordinate tree partitions derived from a level partition. Dashed grey lines indicate
    the initial level partition.}
    \label{fig:coord-tree-full-parts}
\end{figure}

\begin{table*}[ht]
    \centering
    \begin{tabular}{|c|ll|}
        \hline
        Level Type &  \multicolumn{2}{c|}{Level Function Definitions} \\
        \hline
        \multirow{2}{*}{\dense{}} & 
\begin{lstlisting}
initUniversePartition(): return `C = {}`
createUniversePartitionEntry(color, bounds): 
  return `C[color] = bounds`
finalizeUniversePartition(): 
  return `P = partitionByBounds(C, dom)`, `P`, `P`
\end{lstlisting} & 
\begin{lstlisting}
initNonZeroPartition(): return `C = {}`
createNonZeroPartitionEntry(color, bounds): 
  return `C[color] = bounds`
finalizeNonZeroPartition(): 
  return `P = partitionByBounds(C, dom)`, `P`, `P`
\end{lstlisting} \\
        \cline{2-3}
        & 
\begin{lstlisting}
partitionFromParent(parentPart):
  return `part = copy(parentPart)`, `part`
\end{lstlisting} & 
\begin{lstlisting}
partitionFromChild(childPart):
  return `part = copy(childPart)`, `part`
\end{lstlisting} \\
        \hline
        \multirow{2}{*}{\compressed{}} & 
\begin{lstlisting}
initUniversePartition(): return `C_crd = {}`
createUniversePartitionEntry(color, bounds):
  return `C_crd[color] = bounds`
finalizeUniversePartition():
  return `P_crd = partitionByValueRanges(C_crd, crd)
             P_pos = preimage(pos, P_crd, crd)`, 
             `P_pos`, `P_crd`
\end{lstlisting} & 
\begin{lstlisting}
initNonZeroPartition(): return `C_crd = {}`
createNonZeroPartitionEntry(color, coordBounds):
  return `C_crd[color] = coordBounds`
finalizeNonZeroPartition():
  return `P_crd = partitionByBounds(C_crd, crd)
             P_pos = preimage(pos, P_crd, crd)`, 
             `P_pos`, `P_crd`
\end{lstlisting} \\
        \cline{2-3}
        & 
\begin{lstlisting}
partitionFromParent(parentPart):
  return `P_pos = copy(parentPart)
             P_crd = image(pos, P_pos, crd)`, `P_crd`
\end{lstlisting} & 
\begin{lstlisting}
partitionFromChild(childPart):
  return `P_crd = copy(childPart)
             P_pos = preimage(pos, P_crd, crd)`, `P_pos`
\end{lstlisting} \\
        \hline
    \end{tabular}
    \caption{Partitioning level function definitions for \dense{} and \compressed{} levels. Ticks indicate IR fragments.}
    \label{tab:level-function-impls}
\end{table*}

\subsection{Intuition}\label{sec:intuition}

To build intuition for our approach, we appeal to the coordinate
tree representation of a sparse tensor.
Consider the row-based and non-zero-based SpMV algorithms
discussed in \autoref{sec:programming-model:spmv}.
The distributed loops correspond to 
partitions of the matrix's coordinate tree.
The first strategy distributes the rows, corresponding to a partition
of the first coordinate tree level (\autoref{fig:coord-tree-univ-full-part}).
The second strategy distributes the non-zero coordinates, corresponding
to a partition of the second coordinate tree level (\autoref{fig:coord-tree-full-coord-part}).
In the row-based strategy, each coordinate in the first level will access all 
child coordinates in the second level of the tree.
In the non-zero based strategy, each coordinate in the second level
needs to access its parent coordinate in the first level of the tree.

\autoref{fig:coord-tree-univ-full-part} and \autoref{fig:coord-tree-full-coord-part} depict
the full coordinate tree partitions for each strategy.
Given a partition of a coordinate tree level, we obtain a partition
of the child level by applying the partition to the children of each node in the level.
Conversely, we obtain a partition of the parent level of a partitioned
level by partitioning the parent level such that each parent node is colored with all 
of its children's colors.
\EDIT{
The resulting partition of the coordinate tree may assign a node in a 
multiple colors.
For example, in \autoref{fig:coord-tree-full-coord-part}, the coordinate 0 in
the first level is needed by nodes colored red and green in the second level.
}

\EDIT{
This intuition yields a high level code generation strategy:
First, we create an initial partition of a level of each tensors' coordinate tree
based on the data or computation distribution directives.
Then, we use the initial level partition to create partitions of all levels
above and below the initial level.
}

\subsection{Format Abstractions for Sparse Tensor Partitioning}

\EDIT{To realize the intuitive algorithm on coordinate trees, we must 
translate the partitioning operations on coordinate tree levels to
partitioning operations on the abstract data structures that encode
distributed sparse tensors.}
In \name{}, each tensor dimension is encoded by a level format that encodes how
a coordinate tree level is stored in memory.
Chou et al.~\cite{taco_formats} proposed a compile-time interface for level formats
that provides an abstraction for a code generation algorithm to target.
The abstraction allows for per-dimension reasoning about sparse tensors 
and for new formats implementing the interface to be added without
changing the code generation algorithm.
The format abstraction contains \emph{level functions} that return 
intermediate representation (IR) fragments for the code generator
to manipulate.
In this section, we introduce new level functions
that correspond to the two intuitive phases of coordinate tree partitioning: 
1) initially partitioning a coordinate tree level and 2) deriving a partition 
of the full coordinate tree.
Then, in \autoref{sec:codegen} we show how a code generation algorithm
can utilize these abstractions to generate specialized partitioning code.

There are two groups of functions for creating initial level partitions,
corresponding to universe and non-zero partitions.
The universe partition group consists of three functions:
\begin{lstlisting}
initUniversePartition() -> IRStmt
createUniversePartitionEntry(color,bounds) -> IRStmt
finalizeUniversePartition() -> (IRStmt,partition,partition)
\end{lstlisting}
\code{initUniversePartition} initializes any necessary data structures for partitioning.
\code{createUniversePartitionEntry} takes symbolic values of a color and a
coordinate range that should be assigned to the color and maps
the range of coordinates to the color.
\code{finalizeUniversePartition} finalizes any data structures
and returns a partition to use for partitioning parent levels
and a partition to use for partitioning child levels.
%

The non-zero partition group also consists of three functions:
\begin{lstlisting}
initNonZeroPartition() -> IRStmt
createNonZeroPartitionEntry(color,bounds) -> IRStmt
finalizeNonZeroPartition() -> (IRStmt,partition,partition)
\end{lstlisting}
\code{initNonZeroPartition} and \code{finalizeNonZeroPartition} are the same
as their universe partition counterparts.
\code{createNonZeroPartitionEntry} is similar to \code{createUniversePartitionEntry},
but takes bounds on the positions within the level that encode non-zero coordinates.
%

There are two functions for constructing derived partitions:
\begin{lstlisting}
partitionFromParent(partition) -> (IRStmt,partition)
partitionFromChild(partition) -> (IRStmt,partition)
\end{lstlisting}
Each function partitions a level using an existing partition,
and returns a partition to use to partitioning child or parent levels.

We show implementations of the partition level functions for \dense{}
and \compressed{} levels in \autoref{tab:level-function-impls}.
The initial partitioning functions for \dense{} levels color the coordinate
space that the \dense{} level represents, and the derived partitioning functions
apply input partitions to the coordinate space.
For \compressed{} levels, the universe partitioning functions partition the
\crd{} region by bucketing the coordinates into the ranges demarcated
by \code{createUniversePartitionEntry}, while the 
non-zero partitioning functions partition the \crd{} region
according to the position bounds demarcated by \code{createNonZeroPartitionEntry}.
Both use a preimage to recover a partition of the \pos{} region.
The derived partitioning functions partition the \pos{} and \crd{} regions
from parent and child partitions, and use image and preimage to create
partitions of the other region in the level.

\subsection{Code Generation Algorithm}\label{sec:codegen}

We describe how the format abstraction functions for partitioning are used in an
algorithm to generate code that partitions sparse tensors.
\EDIT{
This algorithm implements the intuitive strategy described in \autoref{sec:intuition}
by making calls to the format abstraction functions to generate code that
partitions each level of a tensor.
Our algorithm encodes information about the relationships between
tensor levels through creating partitions, and discharges the data movement
operations required to materialize these partitions to a runtime system.
}

\begin{figure}[!ht]
\begin{subfigure}{\linewidth}
\begin{lstlisting}[frame=none,language=Python,commentstyle=\color{gray},numbers=left, xrightmargin=0.1cm,xleftmargin=0.4cm]
def codegen(TINStatement s, IndexVar i):
  if not distributed(i):
    # Fall back to standard TACO code generation.
    codegenTACO(s, i)
    return
    
  if coordinateValueIteration(s, i):
    # Create initial partitions of each tensor.
    createInitialUniversePartitions(i, s)
    # Derive full coordinate tree partitions.
    partitionCoordinateTrees(i, s)
  else:
    # Create initial partition of non-zero split tensor.
    createInitialNonZeroPartition(i, s)
    # Partition the full coordinate tree of the 
    # non-zero split tensor.
    partitionNonZeroCoordinateTree(i, s)
    # Using the partition of the non-zero split tensor, 
    # partition all other accessed tensors.
    partitionRemainingCoordinateTrees(i, s)
  
  # Emit a distributed for loop over the index variable 
  # and pass the partitions to each iteration.
  emitDistributedForLoop(i)
  # Codegen the next index variable as the loop body.
  codegen(s, next(i))
\end{lstlisting}
    \caption{Code generation algorithm.}
    \label{fig:codegen-alg}
\end{subfigure}

\begin{subfigure}{\linewidth}
    \begin{lstlisting}[frame=none,language=C++,commentstyle=\color{gray},numbers=left, xleftmargin=0.4cm,xrightmargin=0.1cm,keywords={void, for, int,auto, distributed},mathescape]
void SpMV(Tensor a, Tensor B, Tensor c, int pieces) {
  // B1.initUniversePartition()$\tikzmark{initialUniversePartStart}$
  Coloring BColoring = {};
  for (int io = 0; io < pieces; io++) {
    int iLo = io * (B[0].dim / pieces);
    int iHi = (io + 1) * (B[0].dim / pieces);
    // B1.createUniversePartitionEntry(io, {iLo, iHi})
    BColoring[io] = {iLo, iHi - 1}; }
  // B1.finalizeUniversePartition()
  auto B1Part = partitionByBounds(B[0].dom, BColoring);$\tikzmark{initialUniversePartEnd}$
  // B2.partitionFromParent(B1Part)$\tikzmark{coordTreeStart}$
  auto B2PosPart = copy(B1Part, B[1].pos);
  auto B2CrdPart = image(B2PosPart, B[1].pos);
  auto BValsPart = copy(B2CrdPart, B.vals);$\tikzmark{coordTreeEnd}$
  // Execute each iteration on a different node.$\tikzmark{distforstart}$
  distributed for io in {0 ... pieces} {
    B = Tensor({B1Part[io], 
         {B2PosPart[io], B2CrdPart[io]}, 
          BValsPart[io]});$\tikzmark{distforend}$
    for (int ii = 0; ii < (B[0].dim / pieces); ii++) {$\tikzmark{codegenstart}$
      int i = io * (B[0].dim / pieces) + ii;
        for (int jB = B[1].pos[i].lo; 
                    jB <= B[1].pos[i].hi; jB++) {
          int j = B[1].crd[jB];
          a.vals[i] += B.vals[jB] * c.vals[j];
        }}}}$\tikzmark{codegenend}$
\end{lstlisting}
    \caption{Generated pseudo-code for a row-based SpMV. 
    For simplicity, partitioning of $a$ and $c$, iteration guards 
    and bounds checks are omitted.}
    \label{fig:example-gen-code}
\end{subfigure}

\begin{tikzpicture}[overlay, remember picture]
    \drawBrace[-0.1em]{initialUniversePartStart}{initialUniversePartEnd}{(1)};
    \drawBrace[5.0em]{coordTreeStart}{coordTreeEnd}{(2)};
    \drawBrace[2.9em]{distforstart}{distforend}{(3)};
    \drawBrace[0.6em]{codegenstart}{codegenend}{(4)};
\end{tikzpicture}
    
\begin{subfigure}{\linewidth}
    \begin{subfigure}{0.49\linewidth}
        \centering
        \includegraphics[width=0.9\linewidth]{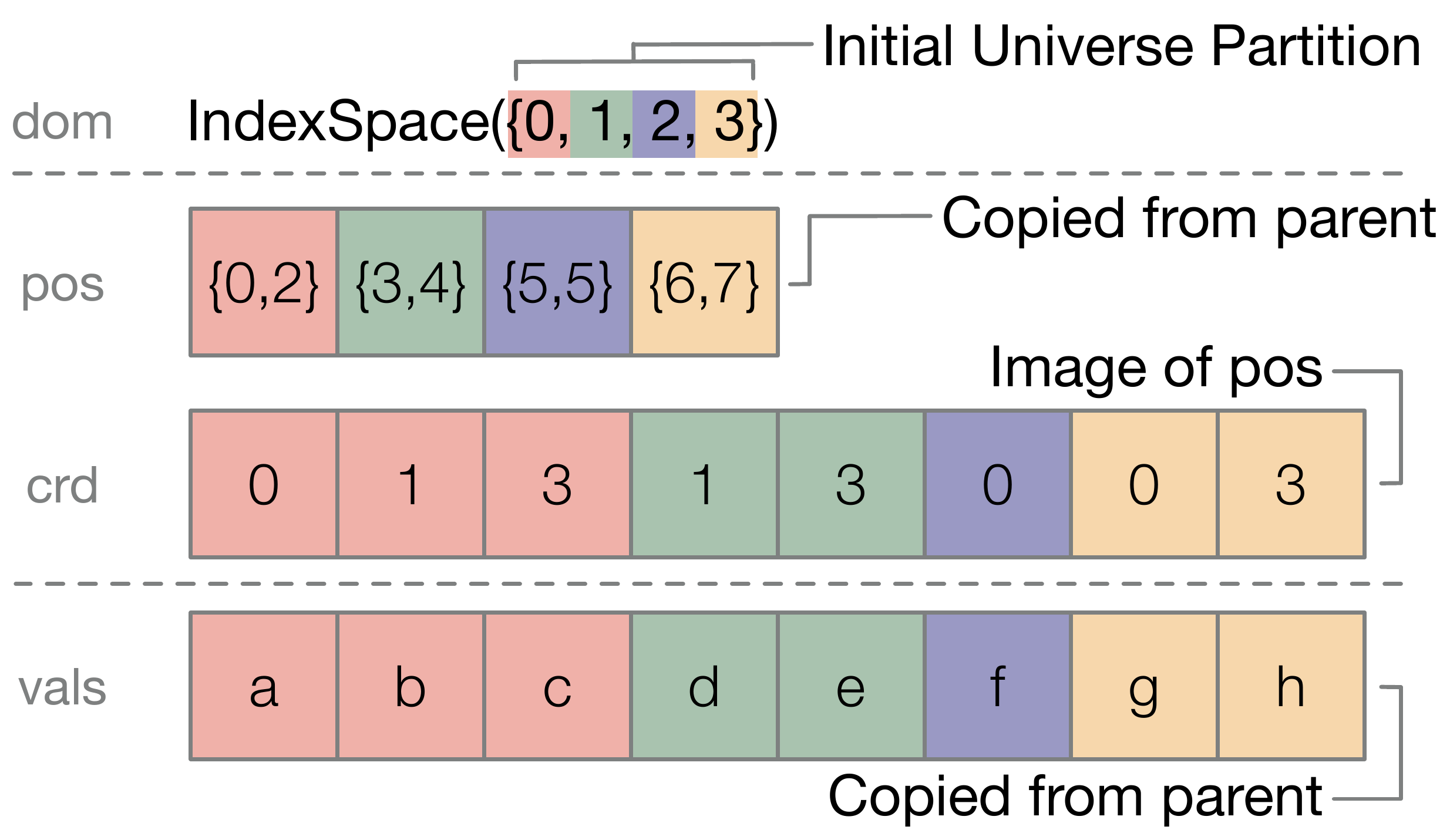}
        \caption{Generated universe partition for SpMV.}
        \label{fig:example-gen-part-univ}
    \end{subfigure}\hfill
    \begin{subfigure}{0.49\linewidth}
        \centering
        \includegraphics[width=0.9\linewidth]{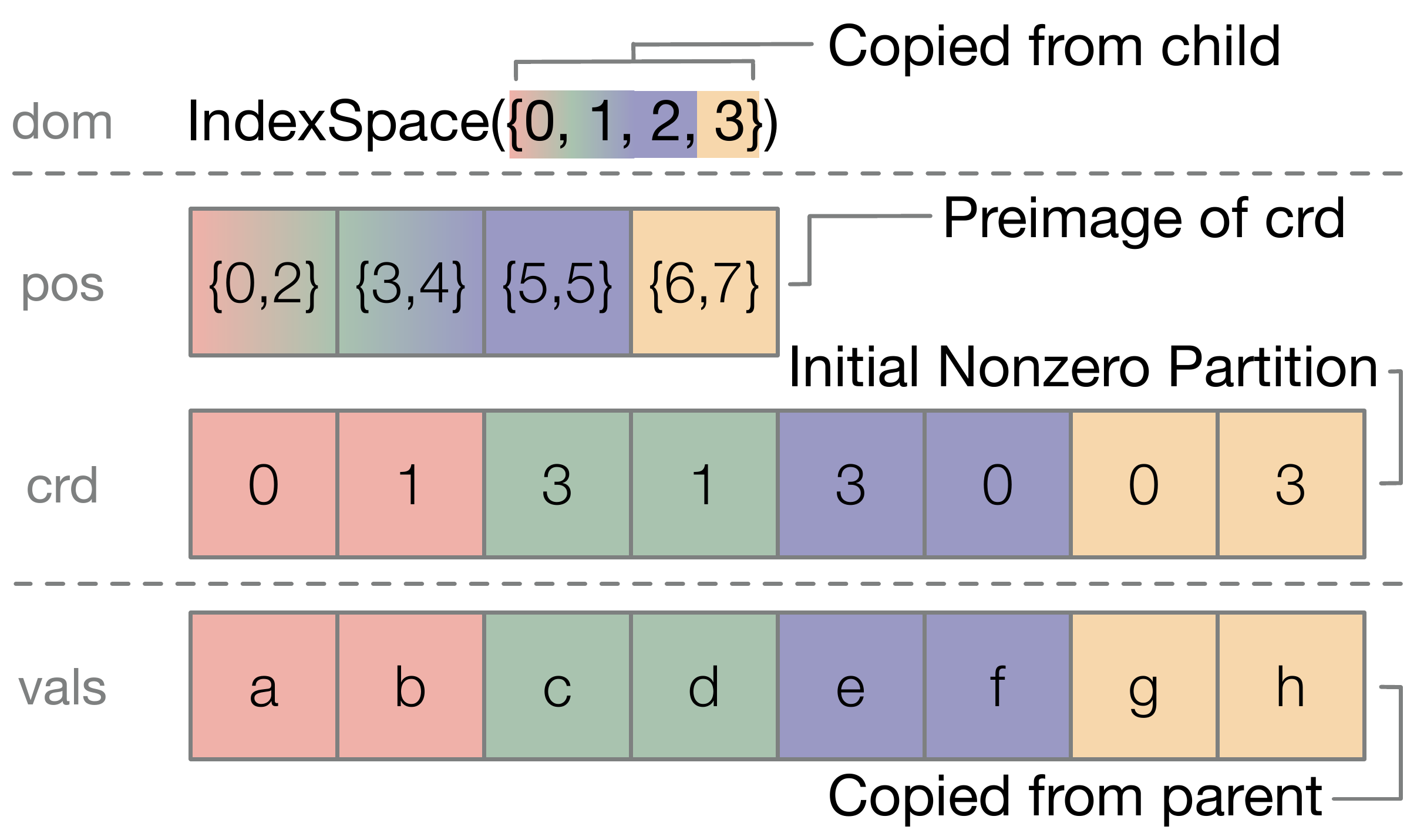}
        \caption{Generated non-zero partition for SpMV.}
        \label{fig:example-gen-part-coord}
    \end{subfigure}
\end{subfigure}
\caption{Code generation algorithm and examples of output.}
\label{fig:algorithm}
\end{figure}

The algorithm is a recursive function called on a scheduled
TIN statement and recurses on index variables in the scheduled order.
For each distributed index variable, the algorithm generates code to create 
initial level partitions of tensors, and then code to derive partitions 
of full coordinate trees.
Pseudo-code for the algorithm is in \autoref{fig:codegen-alg}.
\autoref{fig:example-gen-code} contains generated code
for a row-based SpMV schedule.
\autoref{fig:example-gen-part-univ} and \autoref{fig:example-gen-part-coord}
depict partitions created
by our algorithm for the row-based and non-zero-based SpMV schedules.

During code generation, TACO breaks iteration over sparse data structures into 
two kinds: \emph{coordinate value} iteration and \emph{coordinate position} iteration.
Coordinate value loops iterate over all possible coordinate values and 
compute or retrieve present coordinates from tensor levels.
Coordinate position loops instead iterate over the non-zero coordinates
in a level by directly iterating over the positions that hold non-zero
coordinates in a level.
\EDIT{
Coordinate position loops arise when applied scheduling transformations 
strip-mine iterations over non-zero coordinates only, as discussed in
\autoref{sec:scheduling-language}.
}
Our code generation algorithm follows these two cases:
distributed coordinate value loops distribute the space of coordinates,
corresponding to universe partitions, while
distributed coordinate position loops distribute the space of non-zero
coordinates, corresponding to non-zero partitions.

To generate partitioning code for coordinate value loops, the algorithm
creates initial universe partitions of accessed tensor levels by calling
\code{createInitialUniversePartitions}, which proceeds in three steps \EDIT{ to
generate the code labeled (1) in \autoref{fig:example-gen-code}}.
First, it invokes \code{initUniversePartition} for each tensor accessed
by the current index variable.
Next, it emits a \code{for} loop over the current index variable, and
infers symbolic upper and lower bounds on the index variable to pass to
an invocation of \code{createUniversePartitionEntry} for each tensor.
Lastly, it calls \code{finalizeUniversePartition} on each tensor.
The algorithm then partitions each coordinate tree using
\code{partitionCoordinateTrees} partitions each level above
the initial level with \code{partitionFromChild} and partitions
each level below the initial level with \code{partitionFromParent}.
\EDIT{
The result of \code{partitionCoordinateTrees} is demarcated with
label (2) in \autoref{fig:example-gen-code}.
}

The partitioning process for coordinate position loops is similar
to the process for partitioning coordinate \EDIT{value} loops.
The algorithm calls \code{createInitialNonZeroPartition},
which constructs an initial level partition of the position space
tensor by invoking \code{initNonZeroPartition}, generating a loop with
the result of \code{createNonZeroPartitionEntry} and completing the level
partition with \code{finalizeNonZeroPartition}.
Next, the algorithm partitions the full coordinate tree of the position space tensor
as previously with \code{partitionNonZeroCoordinateTree}.
Finally, the algorithm uses a universe partition derived from the top-level partition
of the position space tensor's coordinate tree to use as an initial
level partition for all other tensors in the statement (\code{partitionRemainingCoordinateTrees}).

After creating the necessary partitions of each tensor in the input statement,
the algorithm emits a distributed for loop over the current index variable
and passes the corresponding sub-region of each partition to each distributed
loop iteration (\code{emitDistributedForLoop})\EDIT{, as seen in label (3) in \autoref{fig:example-gen-code}}.
The body of the loop is generated by a recursive call to the code generation algorithm
\EDIT{(label (4) in \autoref{fig:example-gen-code})}.

\section{Implementation}

We implement \name{} by extending DISTAL~\cite{distal}, which targets the 
Legion~\cite{legion} runtime system, and use techniques from TACO~\cite{taco}
to generate sparse code for CPUs and GPUs.
%

\subsection{Partitioning}
To implement the partitioning strategy described in \autoref{sec:compilation}
on the irregular data structures that store sparse tensors, 
\name{} utilizes the \emph{dependent partitioning}~\cite{deppart} infrastructure of Legion.
We implement the index space, partition and region types discussed 
in \autoref{sec:data-structures}
with Legion's \code{IndexSpace}, \code{Region} and \code{LogicalPartition} types.
Legion supports distributed image and preimage operations,
allowing for partitioning of sparse tensors without
moving all program data into a centralized location.
Legion manages moving the subregions defined by \name{}'s partitions
between memories in the target machine through runtime analysis.

%
%
%

\subsection{Sparse Output Tensors}

Our prototype implementation of \name{} has support for some cases of statements and 
output sparsity formats.
Some tensor index notation statements (such as sparse tensor-times-vector $\textbf{A}(i, j) = \textbf{B}(i, j, k) \cdot c(k)$) preserve
the sparsity pattern of the input tensor in the output tensor.
\name{} identifies situations where this is possible and emits code that copies the coordinate metadata from the
output tensor into the input tensor and modifies the values of the output tensor only.
For cases where the sparsity pattern of the output is unknown, we implement the two-phase parallel assembly approach described
by Chou et al.~\cite{taco_assembly}.
\name{} generates code that first symbolically executes the desired computation and 
records what locations non-zero outputs should be written into, and then uses the 
results of the symbolic execution to construct the output tensor without 
additional synchronization.

\subsection{Tensor Distribution Notation}

We follow DISTAL's~\cite{distal} approach to implement tensor distribution notation.
DISTAL translates a TDN statement into 
a scheduled TIN statement, using \code{divide} 
and \code{distribute} to partition the tensor according to the TDN statement.
We extend this approach with coordinate fusion using
\code{fuse},
and support non-zero partitioning by using the version of 
\code{divide} 
that strip-mines the non-zero coordinates.
The resulting TIN statement is then compiled using the algorithm described
in \autoref{fig:codegen-alg}.

\begin{figure*}[!ht]
    \begin{subfigure}[b]{0.33\textwidth}
        \centering
        \includegraphics[width=0.9\textwidth]{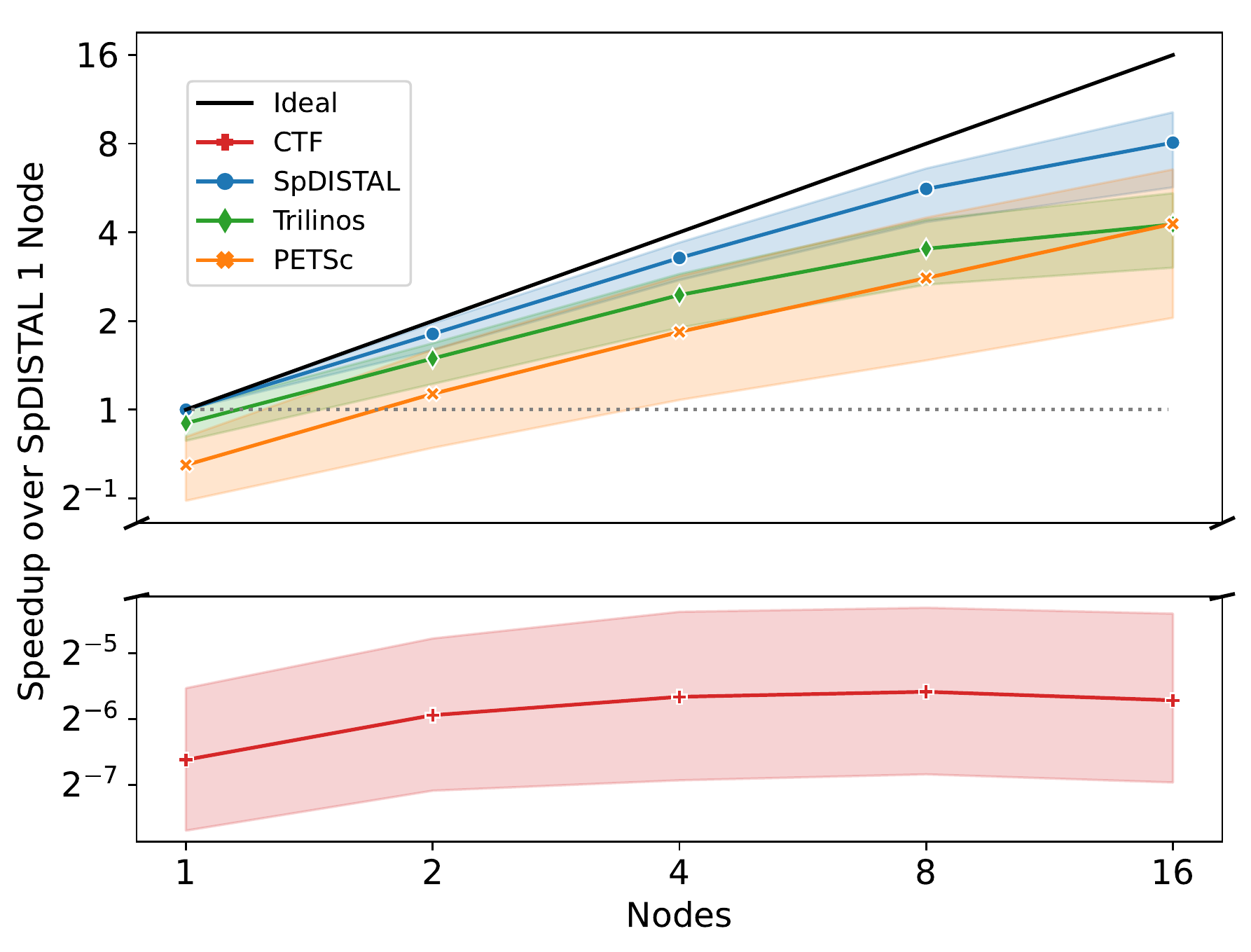}
        \captionof{figure}{SpMV}
        \label{fig:cpu-spmv-strong}
    \end{subfigure}\hfill
    \begin{subfigure}[b]{0.33\textwidth}
        \centering
        \includegraphics[width=0.9\textwidth]{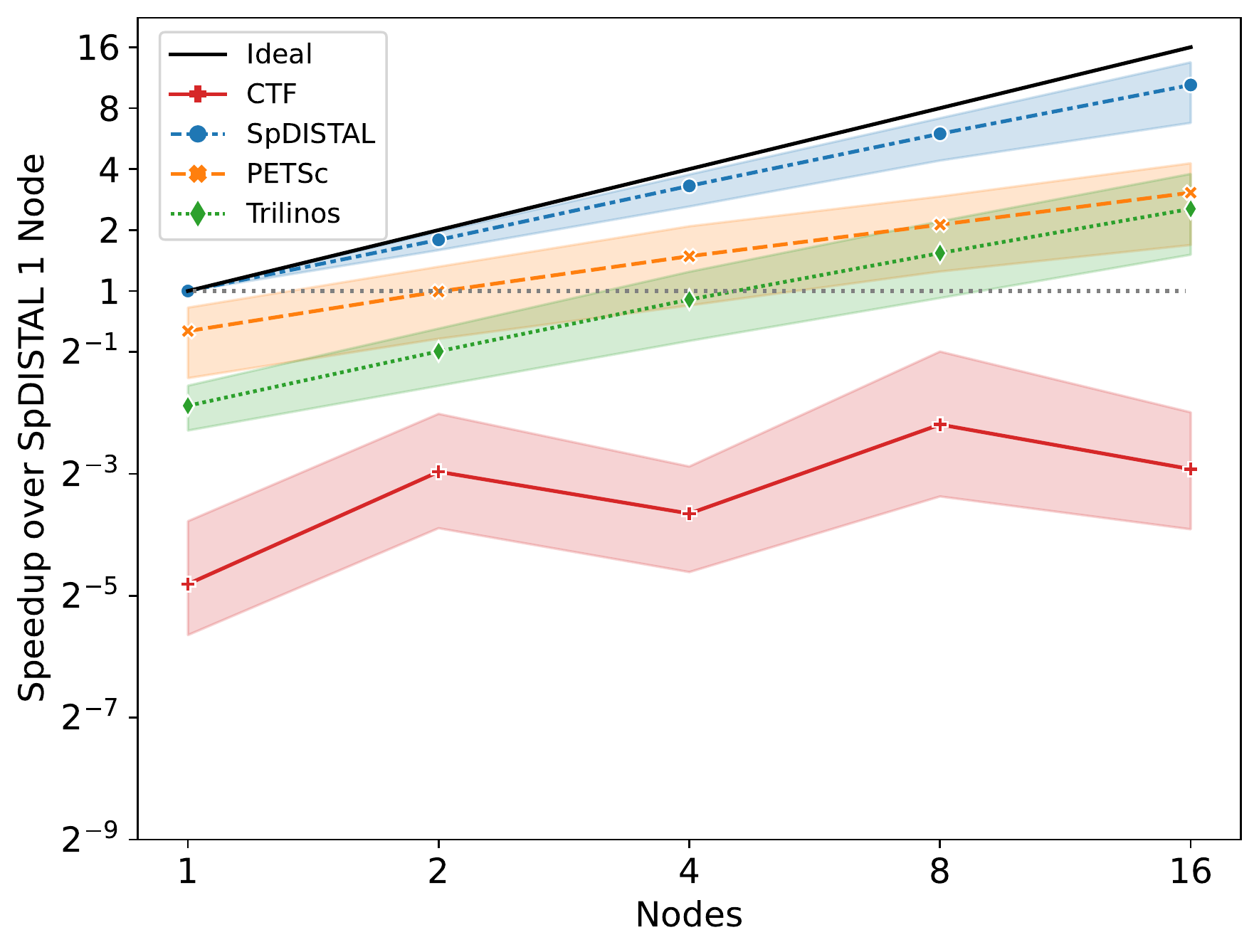}
        \captionof{figure}{SpMM}
        \label{fig:cpu-spmm-strong}
    \end{subfigure}\hfill
    \begin{subfigure}[b]{0.33\textwidth}
        \centering
        \includegraphics[width=0.9\textwidth]{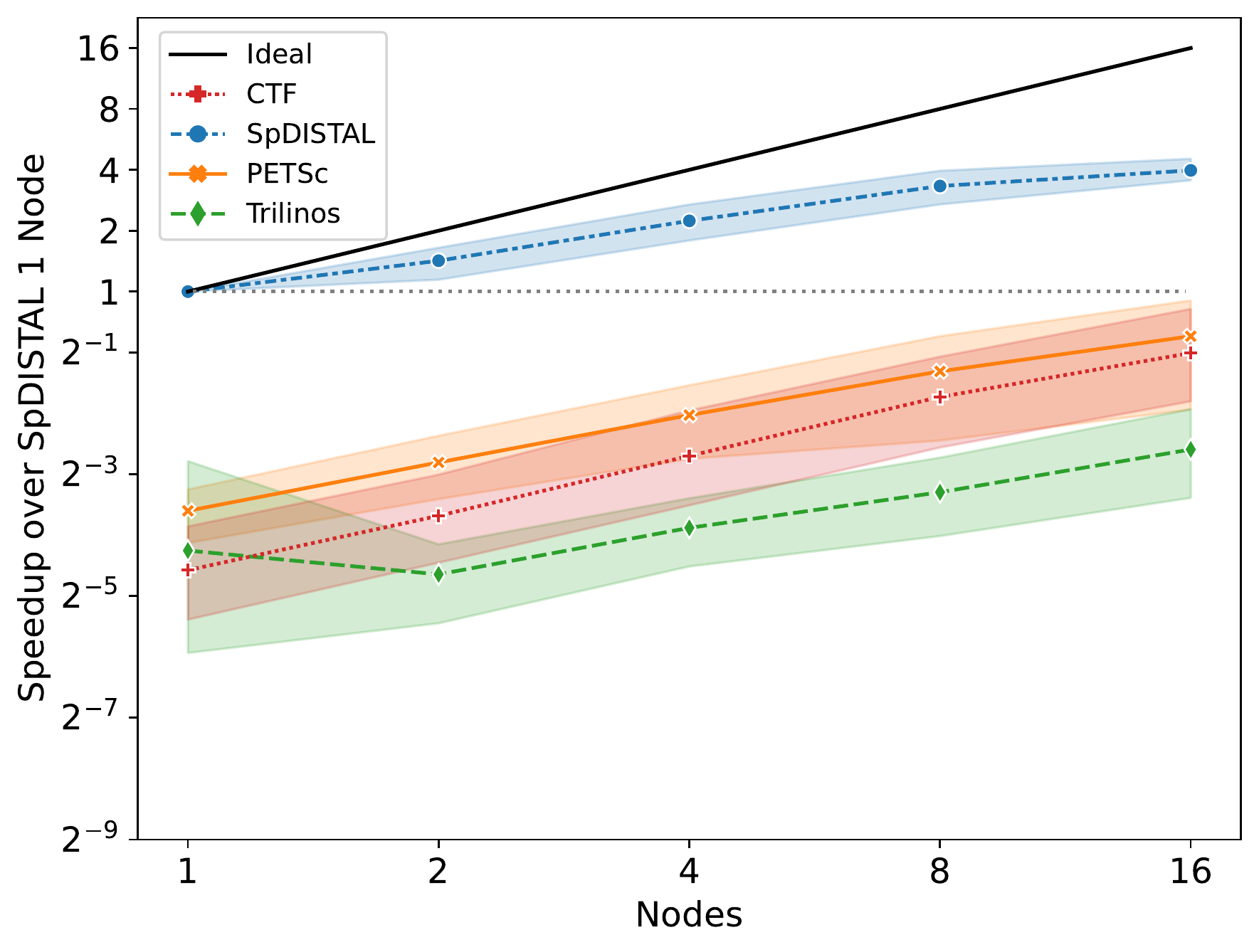}
        \captionof{figure}{SpAdd3}
        \label{fig:cpu-spadd3-strong}
    \end{subfigure}

    \vspace{0.5em}

    \begin{subfigure}[b]{0.33\textwidth}
        \centering
        \includegraphics[width=0.9\textwidth]{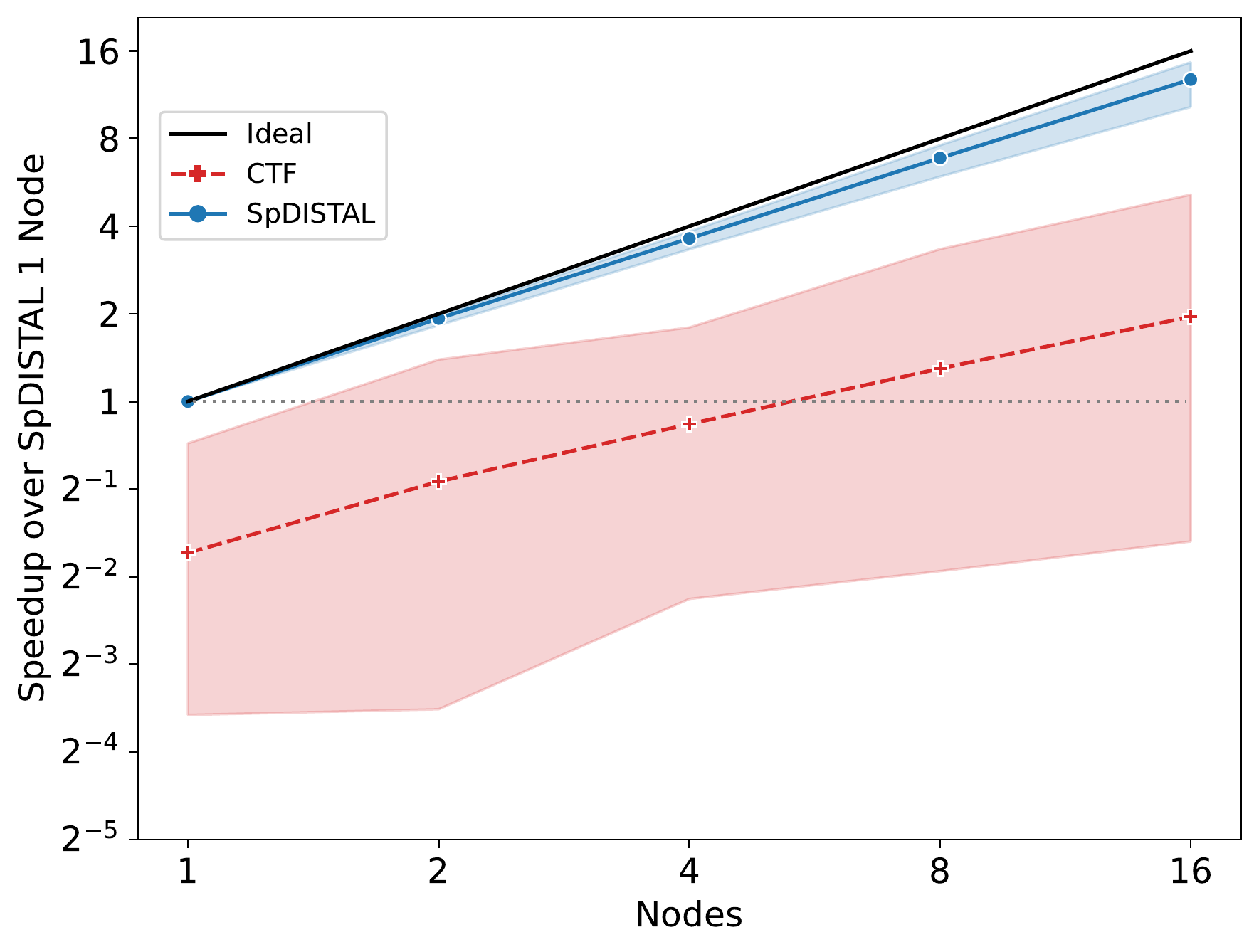}
        \caption{SDDMM}
        \label{fig:cpu-sddmm-strong}
    \end{subfigure}\hfill
    \begin{subfigure}[b]{0.33\textwidth}
        \centering
        \includegraphics[width=0.9\textwidth]{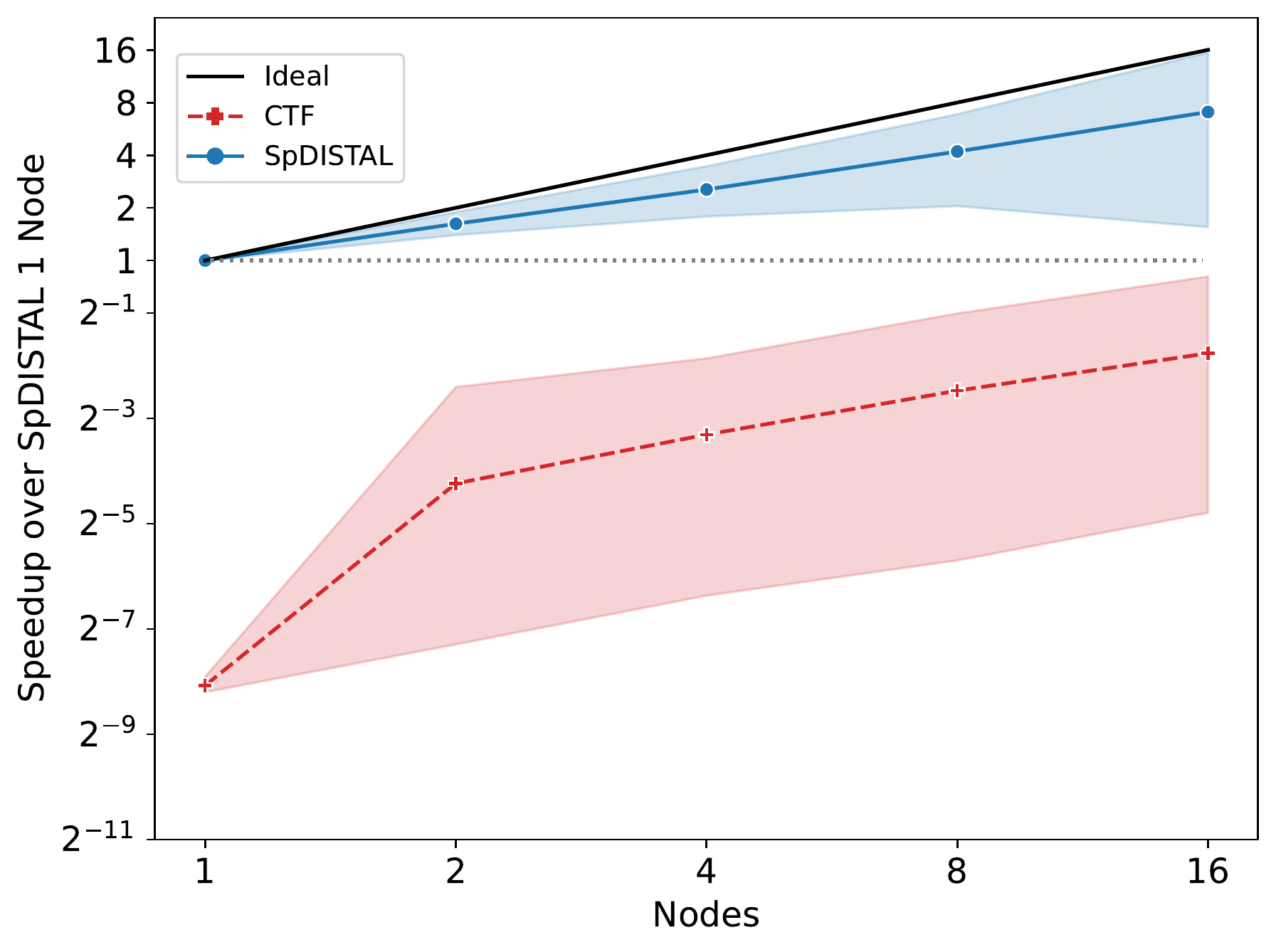}
        \caption{SpTTV}
        \label{fig:cpu-spttv-strong}
    \end{subfigure}\hfill
    \begin{subfigure}[b]{0.33\textwidth}
        \centering
        \includegraphics[width=0.9\textwidth]{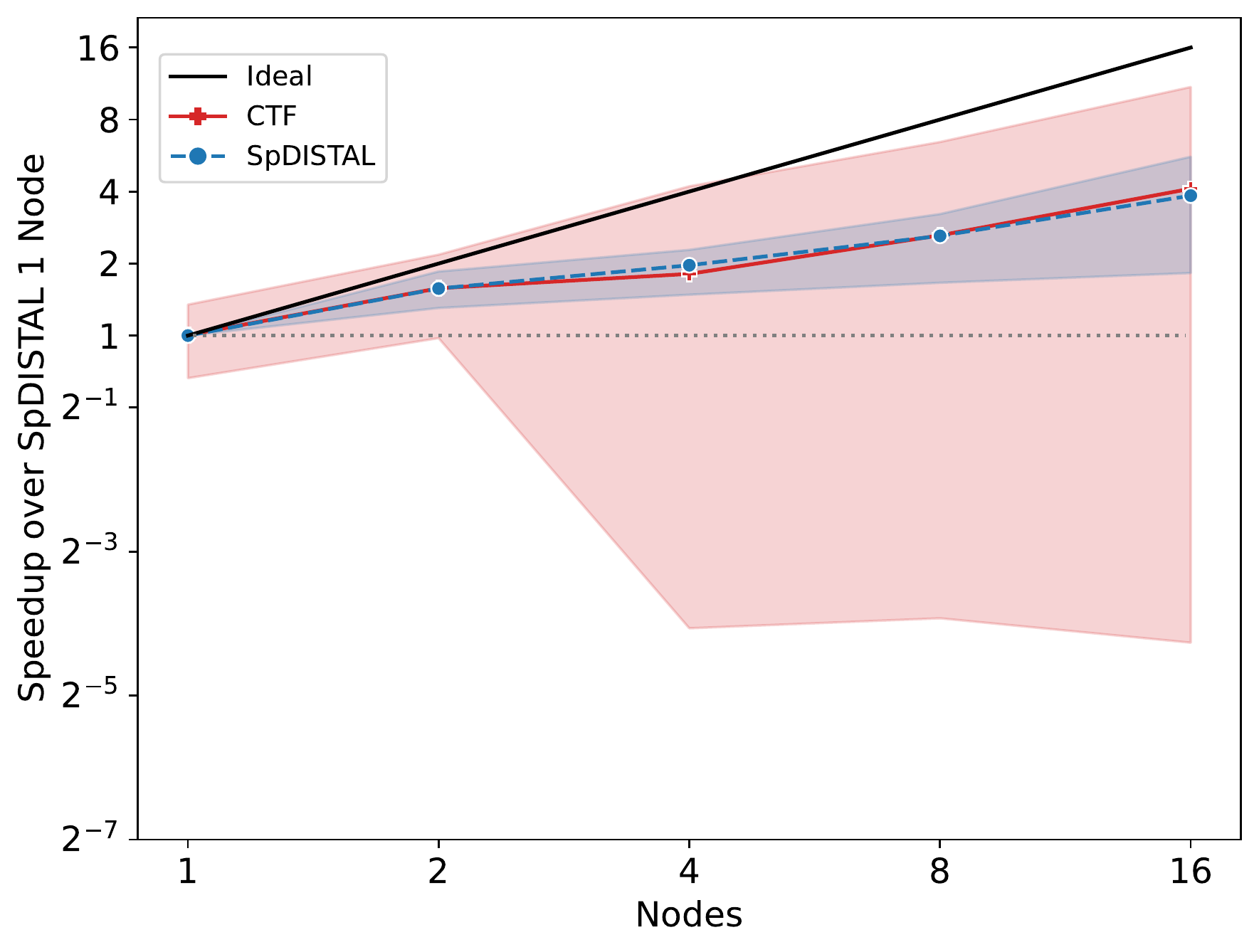}
        \caption{SpMTTKRP}
        \label{fig:cpu-spmttkrp-strong}
    \end{subfigure}
    \caption{Strong scaling results for CPUs. On SpTTV, CTF OOM'ed on the ``patents'' tensor on 1 node. On SpMTTKRP, CTF OOM'ed on the ``freebase\_music'' tensor on 1 and 2 nodes, and on the ``freebase\_sampled'' tensor at all node counts.}
    \label{fig:cpu-strong-scaling}
\end{figure*}

\section{Evaluation}

\emph{Experimental Setup.} 
We ran our experiments on the Lassen supercomputer~\cite{lassen}.
Each Lassen node has a 40 core dual socket IBM Power9,
four NVIDIA Volta V100s connected by NVLink 2.0 and an
Infiniband EDR interconnect.
All systems were compiled with GCC 8.3.1 and CUDA 11.1.
Legion\footnote{\url{https://gitlab.com/StanfordLegion/legion/}, commit excluded for review.} 
was configured with GASNet-EX 2021.3.0 for communication.

\begin{table}[h]
    \centering
    \footnotesize
    \begin{tabular}{|c|c|c|}
        \hline
        Tensor name & Domain & Non-zeros \\
        \hline
        \hline
        arabic-2005 & Web Connectivity & $6.39 \times 10^8$ \\
        \hline
        it-2004 & Web Connectivity & $1.15 \times 10^9$ \\
        \hline
        kmer\_A2a & Protein Structure & $3.60 \times 10^8$ \\
        \hline
        kmer\_V1r & Protein Structure & $4.65 \times 10^8$ \\
        \hline
        mycielskian19 & Synthetic & $9.03 \times 10^8$ \\
        \hline
        nlpkkt240 & PDE's & $7.60 \times 10^8$ \\
        \hline
        sk-2005 & Web Connectivity & $1.94 \times 10^9$ \\
        \hline
        twitter7 & Social Network & $1.46 \times 10^9$ \\
        \hline
        uk-2005 & Web Connectivity & $9.36 \times 10^8$ \\
        \hline
        webbase-2001 & Web Connectivity & $1.01 \times 10^9$ \\
        \hline
        \hline
        freebase\_music & Data Mining & $1.74 \times 10^9$ \\
        \hline
        freebase\_sampled & Data Mining & $9.95 \times 10^7$ \\
        \hline
        nell-2 & NLP & $7.68 \times 10^7$ \\
        \hline
        patents & Data Mining & $3.59 \times 10^9$ \\
        \hline
    \end{tabular}
    \caption{Tensors and matrices considered in our experiments. The first group is matrices from SuiteSparse~\cite{suitesparse}. The second group are 3-tensors, where tensors
    prefixed with ``freebase'' are from the Freebase~\cite{freebase} dataset,
    and remaining tensors are from the FROSTT~\cite{frosttdataset} dataset.}
    \label{fig:experiment-tensors}
\end{table}

\emph{Comparison Targets.} We compare against PETSc\footnote{\url{https://gitlab.com/petsc/petsc}, version 3.16.3, commit e27481de.}, the TPetra~\cite{tpetra-website} package of Trilinos\footnote{\url{https://github.com/trilinos/Trilinos}, version 13.2.0, commit 4a5f7906.}
and  Cyclops Tensor Framework (CTF)\footnote{\url{https://github.com/cyclops-community/ctf}, commit 36b1f6de.}.
Trilinos and CTF were configured with OpenMP.
Trilinos and PETSc were built with CUDA support.

\emph{Dataset.} We consider 14 real-world matrices and tensors from the SuiteSparse~\cite{suitesparse}, FROSTT~\cite{frosttdataset} and Freebase~\cite{freebase} datasets as described in \autoref{fig:experiment-tensors}.
For SuiteSparse, we chose the largest matrices that were representable
in PETSc's and Trilinos's default configuration (32-bit indexing).
For FROSTT, we chose all 3-tensors that satisfy the CTF limitation that tensor dimensions 
must multiply to less than the maximum 64 bit integer.
For operations with multiple sparse inputs, we follow Henry and Hsu et al.~\cite{sparse-array-programming}
by shifting the last dimension of each tensor to construct additional sparse inputs.

\emph{Experimental Methodology.}
All experiments were run with 10 warm-up trials, 20 timed trials and a 90 minute timeout.
For all tensors other than ``patents'' we use a format with a \dense{} outer level and
all other levels \compressed{}; for ``patents'' we use two outer 
\dense{} levels and \compressed{} inner level.
\EDIT{Thus, we use the same compressed format for matrices (CSR) as PETSc and Trilinos.}
\name{}'s kernels were run with one rank per node.
For CPUs, we run PETSc and CTF with one rank per core, and
one rank per socket for Trilinos.
\EDIT{All CPU experiments utilize all cores on each node.}
For GPUs, PETSc and Trilinos were run with one rank per GPU.

\emph{Overview.} We evaluate the performance of \name{}
by comparing against hand-written systems (PETSc and Trilinos),
and interpretation-based systems (CTF), and consider both
strong scaling and weak scaling experiments.
Our evaluation shows that 1) \name{} achieves performance 
competitive with (specialized) expert-tuned kernels in hand-written systems and 2) 
\name{} significantly outperforms (general) interpretation-based approaches
to distributed sparse tensor algebra.
These results indicate that a compilation-based approach that produces
bespoke implementations can achieve both generality and high performance,
and \name{} is the first such approach for distributed sparse tensor algebra.

\subsection{Strong Scaling Performance.}

We evaluate \name{} on the following sparse tensor kernels, all of which have been
used in prior work~\cite{taco, taco_scheduling} to evaluate sparse tensor compilers.
\EDIT{SpMV has applications in scientific computing, SpMM and SDDMM appear in sparse machine learning, and SpTTV and SpMTTKRP are used in tensor factorizations arising in data analytics.
SpAdd3 is a synthetic benchmark intended to show the benefits of kernel fusion over binary kernels commonly used in libraries.
}
%
\begin{itemize}
    \item SpMV: $a(i) = \textbf{B}(i, j) \cdot c(j)$
    \item SpMM: $A(i, j) = \textbf{B}(i, k) \cdot C(k, j)$
    \item SpAdd3: $\textbf{A}(i, j) = \textbf{B}(i, j) + \textbf{C}(i, j) + \textbf{D}(i, j)$
    \item SDDMM: $\textbf{A}(i, j) = \textbf{B}(i, j) \cdot C(i, k) \cdot D(k, j)$
    \item SpTTV: $\textbf{A}(i, j) = \textbf{B}(i, j, k) \cdot c(k)$
    \item SpMTTKRP: $A(i, l) = \textbf{B}(i, j, k) \cdot C(j, l) \cdot D(k, l)$
\end{itemize}

Bolded tensors are sparse while all others are dense.
For CPUs, we compare against PETSc, Trilinos and CTF for SpMV, SpMM and SpAdd3.
The remaining operations are unsupported by PETSc and Trilinos.
For GPUs, we restrict our comparison to PETSc and Trilinos, as CTF does 
not currently have GPU support that we could use\footnote{The CTF developers are in progress 
addressing issues with the GPU backend, but this work
was not completed at the time of writing.}.
Both PETSc and Trilinos have GPU support for the SpMV and SpMM kernels, while only Trilinos
has GPU support for the SpAdd3 kernel when the output non-zero pattern is unknown.
For the remaining higher order expressions we compare against \name{}'s CPU kernel.

\begin{figure*}[!ht]
    \centering
    \includegraphics[width=\textwidth]{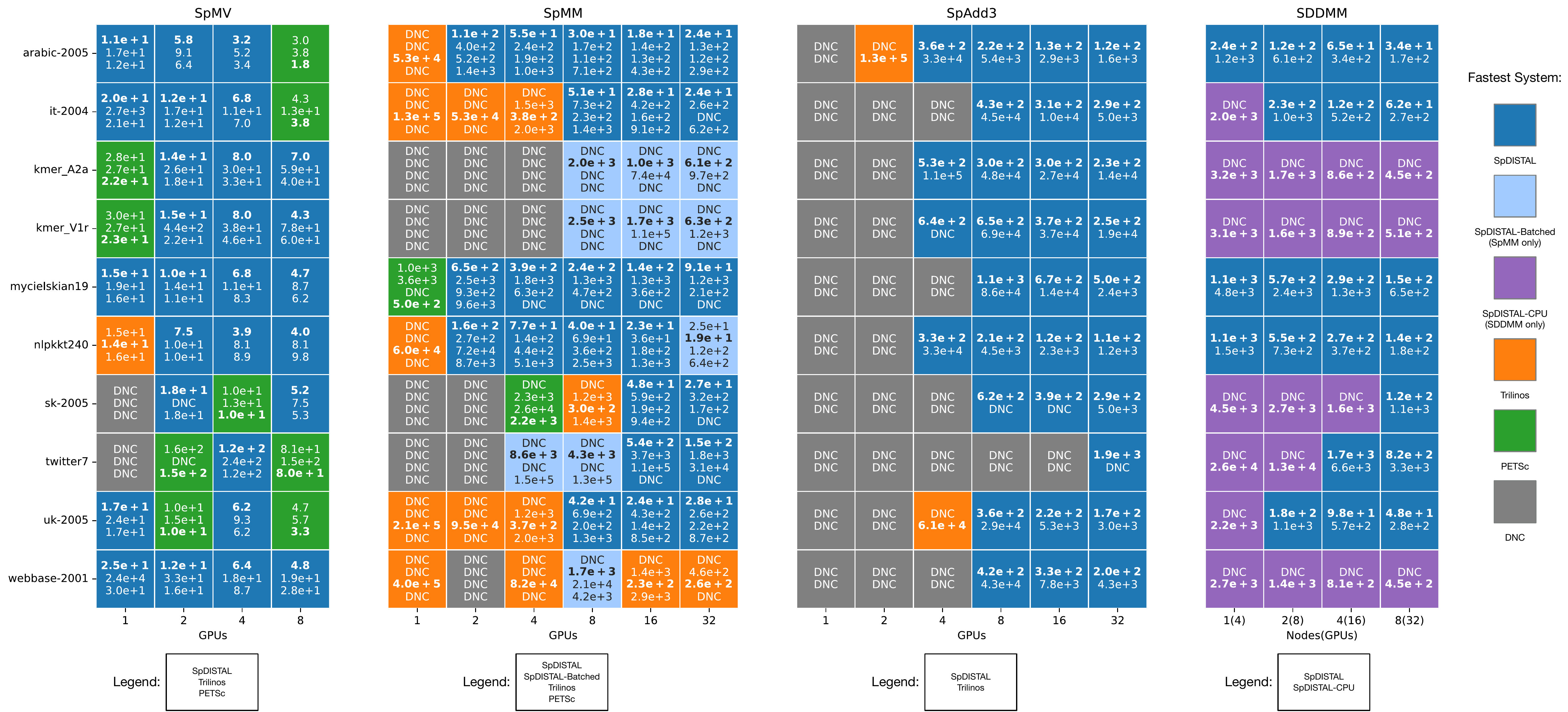}
    \caption{
        GPU strong scaling results for SpMV, SpMM, SpAdd3 and SDDMM. Each box contains the time in milliseconds by each systems' GPU kernels. 
        DNC (does not complete) indicates that a system failed with an error (OOM, timeout, or other). \name{}-Batched is a memory-conserving
        SpMM schedule and \name{}-CPU is \name{}'s CPU SDDMM kernel.
    }
    \label{fig:linalg-gpu}
\end{figure*}

\subsubsection{CPU Results}

\autoref{fig:cpu-strong-scaling} shows speedup plots on CPUs.
Each data point is normalized against \name{} on 1 node,
and each line is the average speedup over all tensors
displayed with a 99\% colored confidence interval (computed over all tensors).
\EDIT{
For all kernels other than SDDMM, we utilize an outer dimension-based 
distributed algorithm and initial data distributions.
For SDDMM, we utilize a non-zero based distributed algorithm and initial data distribution.
We experimented with non-zero based parallelization for SpMV, SpMM, SpTTV and SpMTTKRP
but found that the extra synchronization required within the leaf kernel
costed more than performance gained through load balance.
%
SpAdd3 on CSR matrices is incompatible with the non-zero splitting scheduling transformation, so
we used a row-based distribution and distributed algorithm.
}
%

%
For kernels that PETSc and Trilinos directly support (SpMV and SpMM),
PETSc and Trilinos achieve performance competitive with \name{}.
%
\name{} achieves median speedups of 1.8x and 1.2x on SpMV and 2.01x 
and 3.8x on SpMM over PETSc and Trilinos respectively.
\EDIT{
We attribute the slight performance improvement on SpMV to Legion's deferred 
execution model that avoids unnecessary synchronization.
\name{} achieves a larger speedup over PETSc due to PETSc's current lack of support for
multi-threading: \name{} uses OpenMP to dynamically load balance among threads, yielding an improvement.
For SpMM, we implement the schedule used by Senanayake et al~\cite{taco_scheduling} as the
leaf kernel, which appears to outperform the kernel used by PETSc and Trilinos,
resulting in lower execution time but similar scaling.
}

The benefit of \name{}'s ability to fuse computation into a single kernel is demonstrated by SpAdd3.
PETSc and Trilinos do not implement SpAdd3 and must use two matrix additions.
This approach loses data locality and results in additional
sparse matrix assembly operations, allowing
\name{} to achieve median speedups of 11.8x and 38.5x over PETSc and
Trilinos.
%

Having shown that \name{} can achieve competitive performance with hand-written libraries,
we now compare against CTF's interpretation-based approach.
CTF interprets tensor algebra expressions pair-wise by reducing them to distributed
matrix-matrix multiplication, element-wise, and transposition operations.
As can be seen in the SpMV and SpTTV speedup charts (\autoref{fig:cpu-spmv-strong} and \autoref{fig:cpu-spttv-strong}), interpretation leads to large 
slowdowns---\name{} achieves median speedups of 299x, 161x and 19.2x on SpMV, SpTTV and SpAdd3.
While interpretation leads to large constant-factor slowdowns for binary kernels,
it leads to asymptotic slowdowns for kernels that require fusion, such as
SDDMM and SpMTTKRP~\cite{taco}.
To address this slowdown, the CTF authors have developed hand-written, specialized kernels for 
SDDMM and SpMTTKRP~\cite{cyclops-specialized-kernels}.
For SDDMM, \name{} achieves a median speedup of 15.3x over CTF, and achieves
near perfect speedup due to its load balanced approach.
For SpMTTKRP, \name{} and CTF achieve similar performance (\name{} achieves median 97\% of CTF's performance) and CTF has wider range of performance.
CTF outperforms \name{} on the ``patents'' tensor, and completes the SpMTTKRP
operation on ``patents'' significantly faster than on much smaller tensors.

\subsubsection{GPU Results}

\begin{figure}
    \centering
    \includegraphics[width=\linewidth]{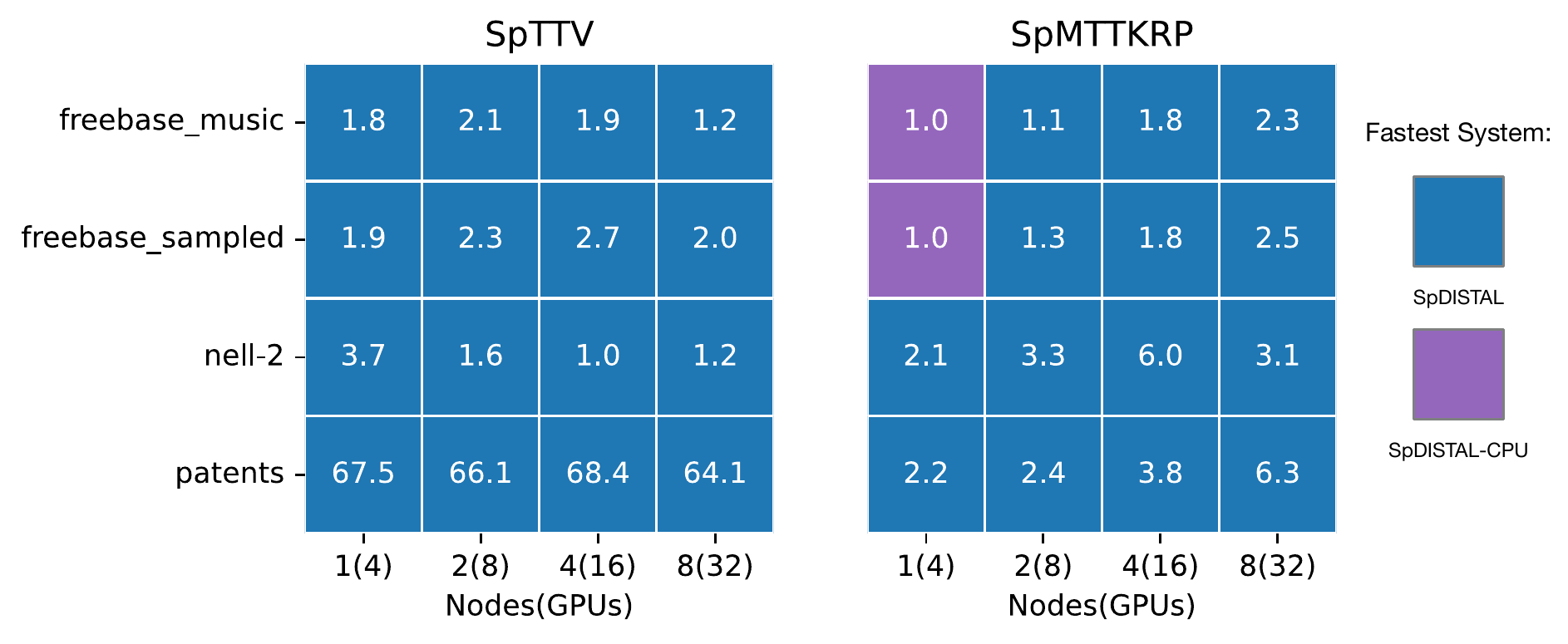}
    
    \caption{GPU strong scaling results for SpTTV and SpMTTKRP comparing \name{}'s GPU and CPU kernels. 
    In each box is the speedup achieved by \EDIT{the faster system over the slower system on the
    same number of nodes}.}
    \label{fig:hot-gpu}
\end{figure}

GPU strong scaling results are shown in \autoref{fig:linalg-gpu} and \autoref{fig:hot-gpu}.
We use a heatmap-based presentation for the GPU results to address that
1) on some kernel and tensor pairs, systems OOM or error out on 
different GPU counts and 2) some kernels have no distributed GPU comparison target.
\EDIT{
These heatmaps display the performance of each system on each input tensor
and processor count pair.
}

\EDIT{We use a row-based algorithm and data distribution for SpMV on GPUs and utilize
CuSPARSE to execute the SpMV at the leaves on a single GPU.
We found that this approach outperformed the distributed version of
the non-zero based schedule utilized by Senanayake et al~\cite{taco_scheduling} 
on the considered dataset and GPU.}
%
%
Due to the short runtime of SpMV on the dataset (order of 10ms), 
we strong scale only to 8 GPUs.
As seen in \autoref{fig:linalg-gpu}, \name{} outperforms PETSc and Trilinos
on 28/38 configurations, and achieves median speedups of 1.07x and 1.65x over PETSc and Trilinos.

SpMM on GPUs had large variability in both the fastest system and 
the number of systems to successfully complete problem instances.
We implement two \name{} schedules for GPU SpMM. 
The first distributes the non-zeros of the computation equally over 
all GPUs \EDIT{at the cost of replicating} the $C$ matrix, leading to OOMs on some 
matrix shapes.
The other schedule (denoted ``\name{}-Batched'') conserves memory by distributing 
the $i$ and $j$ dimensions of the computation to also partition
the $C$ matrix at the cost of potential load imbalance and extra communication.
Per personal communication with the PETSc developers, the PETSc's current
GPU SpMM implementation experiences a significant performance penalty when moving
from one to multiple GPUs.
Trilinos utilizes CUDA-UVM, allowing it fit some problem instances into GPU memory
that \name{} cannot, at the cost of paging this data in and out of the GPU.
As \autoref{fig:linalg-gpu} shows, the \name{}'s load-balanced kernel
performs the best once data fits into memory (24/49 configurations),
and the memory-conserving kernel wins in 10/49 more configurations when
data does not, for a total of 34/49.
There are 13/49 configurations where Trilinos beats \name{}'s memory conserving 
schedule when both fit into GPU memory.
Based on inspection of Trilinos source code, Trilinos performs a single
communication operation to gather the necessary components of the input matrix
to each GPU, and overflowing into CUDA-UVM.
In contrast, \name{}'s memory-conserving algorithm communicates chunks of the dense input
matrix in multiple rounds between nodes to fit data within GPU memory.
We hypothesize that this choice allows for Trilinos to send fewer messages 
over the network than \name{}, leading to faster runtimes on some configurations.

\EDIT{Similar to SpAdd3 on CPUs, we utilize a row-based strategy for SpAdd3 on GPUs.}
PETSc does not support GPU sparse matrix addition when the 
sparsity pattern of the output matrix is unknown,
so we compare against Trilinos for SpAdd3 and find 
that \name{} significantly outperforms Trilinos due to the ability 
to fuse computation and avoid allocation of intermediate results.
\autoref{fig:linalg-gpu} shows that \name{} outperforms Trilinos on 32/34
cases, where Trilinos succeeds in 2/34 cases by fitting matrices into
a smaller GPU count with CUDA-UVM.

For SDDMM, SpTTV and SpMTTKRP we compare \name{}'s 
GPU kernels to \name{}'s CPU kernels using all the resources on a node.
We use a non-zero based \EDIT{algorithm and data} distribution for each kernel.
\EDIT{This differs from the CPU algorithms, as on GPUs,
the additional synchronization within the leaf kernel is outweighed
by the load balance over all GPU threads across the machine.
}
\autoref{fig:linalg-gpu} and \autoref{fig:hot-gpu} show that
\name{}'s GPU kernels achieve a median 4.3x speedup for SDDMM,
2.0x speedup for SpTTV and 2.2x speedup for SpMTTKRP when data 
fits into GPU memory.
On SpMTTKRP, \name{}'s GPU kernels achieve increasing speedup
due to the better load balance offered by the GPU schedule.

\subsection{Weak Scaling Performance.}
\begin{figure}
    \centering
    \includegraphics[width=0.8\linewidth]{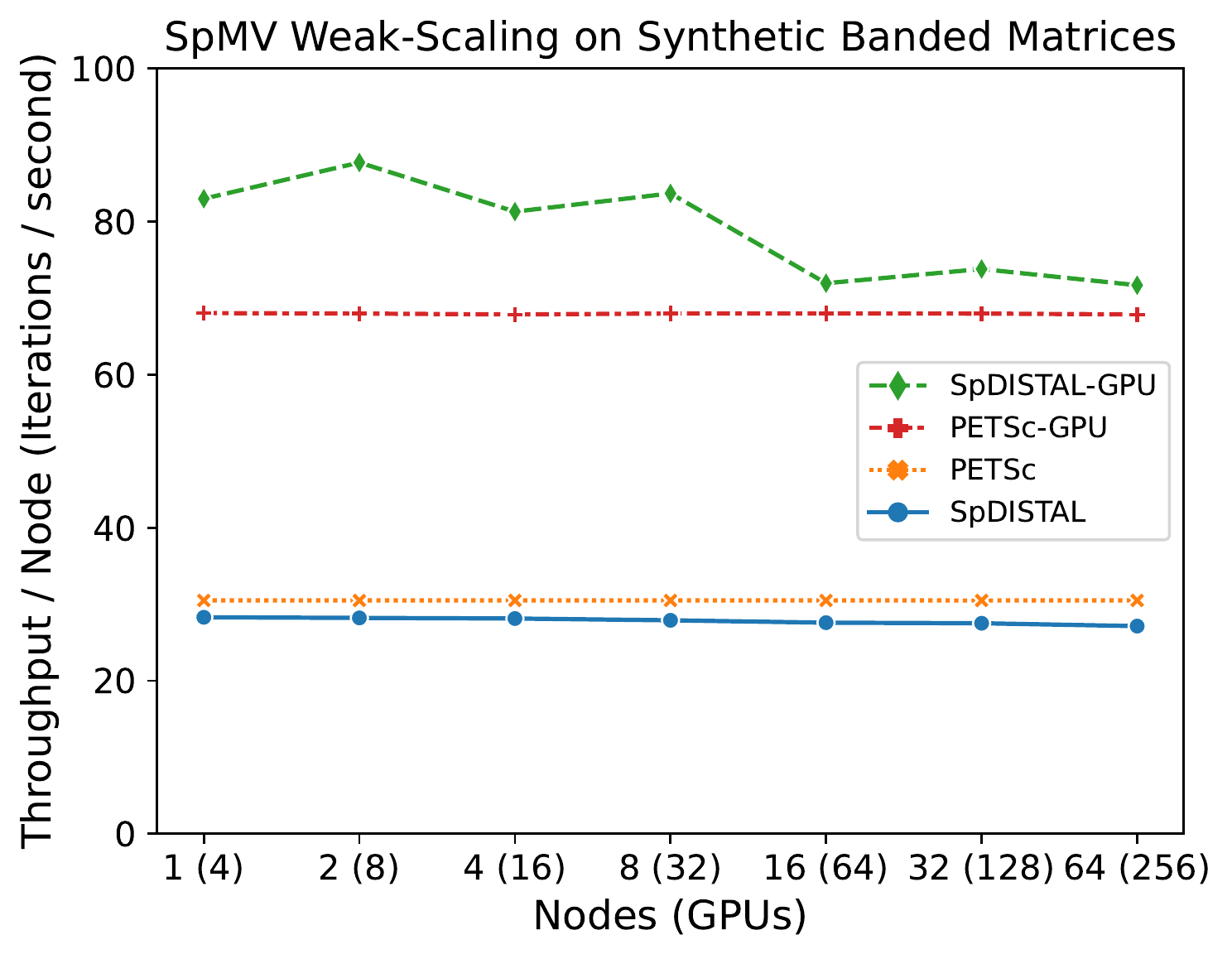}
    \caption{SpMV weak scaling results.}
    \label{fig:spmv-weak-scaling}
\end{figure}

%
\autoref{fig:spmv-weak-scaling} shows the weak-scaling performance of \name{}'s 
SpMV kernel on synthetic banded matrices up to 64 nodes (256 GPUs).
\EDIT{The initial problem size for a single node of CPUs and a single GPU was 700 million non-zeros.}
We compare against PETSc, and exclude Trilinos due to difficulties reading
large matrices from disk.
%
\name{} is configured to use CuSPARSE for local SpMV computations.
We find that PETSc achieves perfect weak-scaling performance
on both CPUs and GPUs.
\name{}'s CPU kernels achieves between 90\% and 92\% of PETSc's performance.
\name{}'s GPU kernel achieves slightly higher performance, between 1.05x and 1.29x, than
PETSc's GPU implementation and has some more performance variability
\EDIT{due to millisecond variations in execution time caused by network effects}.
We attribute the slight improvement to Legion's asynchronous execution model.

\subsection{Discussion}

The scheduling and data distribution languages of \name{} enable concise descriptions of distributed 
algorithms for sparse tensor computations that achieve high performance across
the full range of tensor algebra expressions.
Without scheduling and data distribution languages, users would not be able to describe how
their data and computation map onto distributed machines.

Our experiments show many instances of \name{} outperforming state-of-the-art systems,
enabled by the chosen schedules and data distributions.
For some computations, we replicated algorithms used by existing systems (e.g.
a row-based distribution for SpMV), resulting in \name{} at least equalling
the performance of those systems.
In cases where \name{} exceeded the performance of existing systems using the same
algorithm, we attribute the performance improvement to the efficiency and optimizations
of the Legion runtime.

In some cases, we used \name{}'s capabilities to adapt known algorithms
to new tensor expressions, such as SpADD3.
Our schedule for SpAdd3 extends the row-based algorithm for adding two sparse matrices
to a row-based algorithm that fuses the addition across the three input sparse matrices.
This fusion yields speedups over other systems that perform pair-wise additions
and allocate temporary results, as seen in prior works~\cite{taco, sparse-array-programming}.

Finally, scheduling and data distribution primitives for non-zero partitioning allow
\name{} to express algorithms not found in existing systems.
For SDDMM, GPU SpMM, GPU SpTTV and GPU SpMTTKRP using these primitives yields algorithms
that are statically load-balanced across all processors.
The perfect load balancing enables high performance at scale, regardless of
the input tensor's sparsity structure.

Finally, \name{}'s compiler for the scheduling and data distribution languages 
allows specializing an implementation to a particular sparse tensor computation, 
in contrast to interpretation.
\autoref{fig:cpu-strong-scaling} shows that specialized systems can deliver large speedups
over interpreted approaches on specific kernels.
Significant overheads of interpretation, up to an order of magnitude, have been observed
in prior work~\cite{distal, cyclops-specialized-kernels}.
\name{} provides generality without sacrificing performance.

\section{Related Work}

%

\EDIT{
\name{} is a compiler that implements all of  tensor algebra for sparse tensors and targets distributed machines.
Prior work includes libraries that implement subsets of sparse tensor
algebra on distributed machines, dense compilers that can target single node and
distributed systems, and sparse compilers that target single node systems.

\name{} and Cyclops Tensor Framework~\cite{ctf-main, ctf-sparse} (CTF) are the only two 
distributed systems we know of that support all of tensor algebra on 
sparse tensors.
CTF is an interpreter for tensor algebra.
We show in our evaluation that \name{} significantly outperforms CTF and achieves this
performance through novel compilation techniques that specialize the generated code 
to the input computation, data layout, data distribution and computation distribution.

PETSc~\cite{petsc-web-page} and Trilinos~\cite{trilinos-website} are distributed
sparse linear algebra libraries that support different sparse matrix data structures
and algorithms for sparse linear algebra.
These systems achieve high performance for the subset of tensor algebra implemented within the library, but often have sub-optimal performance for computations that compose multiple library functions.
Our evaluation shows that \name{} can achieve performance competitive with
these specialized systems on computations that they natively implement.
Herault et al.~\cite{block-sparse-contract} develop a multi-GPU system for binary
contractions between block-sparse tensors, but do not support all of tensor
algebra or sparsity structures that are not block-sparse.

Sparse compilation techniques for single-node systems have seen attention from researchers.
Bik et al.~\cite{MT1}, developed an early compiler
that transformed dense loops over matrices into sparse loops over the non-zero coordinates.
The TACO~\cite{taco} compiler was the first to describe how to compile all of sparse tensor
algebra to CPUs.
Chou et al.~\cite{taco_formats} showed how to extend TACO with an abstraction
for definition of new formats without changing the code generation algorithm.
Kjolstad et al.~\cite{taco_workspaces} and Senanayake et al.~\cite{taco_scheduling}
extend TACO with sparse iteration space transformations and GPU code generation.
Finally, Henry and Hsu et al.~\cite{sparse-array-programming} generalize TACO
beyond addition and multiplication to arbitrary functions.
\name{} sets itself apart from these work by showing how to compile sparse tensor algebra
to distributed machines containing CPUs and GPUs.
However, \name{} utilizes components from these prior works for its programming model and implementation.

Finally, dense compilers for both single-node and distributed systems have been developed.
Single-node dense compilers include Halide~\cite{halide}, TVM~\cite{TVM}, 
Tensor Comprehensions~\cite{tensor-comprehensions} and Tiramisu~\cite{tiramisu}.
Tiramisu and Distributed Halide~\cite{dist-halide} allow for targeting distributed backends.
These compilers express computations in high-level DSLs and optimizing transformations
through scheduling languages, similar to \name{}.
DISTAL~\cite{distal} is a dense tensor algebra compiler that supports separate
specifications of data and computation distribution.
The key difference of \name{} from these works is that \name{} enables
(distributed) computation over sparse data.

}

\section{Conclusion}

We introduce \name{}, a system that for
distributed sparse tensor algebra through
independent specifications of computation,
sparse data structures, data distribution and computation distribution.
\name{} realizes this programming model through novel compiler techniques
for data partitioning, and shows that programming distributed sparse tensor
algebra can be general, high performance and productive.


\section*{Acknowledgements}

We thank our anonymous reviewers for their valuable comments that
helped us improve this manuscript.
We thank Scott Kovach, Elliot Slaughter, Axel Feldmann, Shiv Sundram and Olivia Hsu
for their comments and discussion on early stages of this manuscript.
We thank the Legion team for their feedback and support during the development
of \name{}.
We thank the PETSc, Trilinos and CTF authors for their assistance in setup and benchmarking
of their software.
Rohan Yadav was supported by an NSF Graduate Research Fellowship.
This work was supported in part by the Advanced Simulation and Computing (ASC) program of the US Department of Energy’s National Nuclear Security
Administration (NNSA) via the PSAAP-III Center at Stanford, Grant No. DE-NA0002373 and the Exascale Computing Project (17-SC-20-SC), a collaborative effort of the U.S. Department of Energy Office of Science and the National Nuclear Security Administration; the U.S. Department of Energy, Office of Science under Award DE-SCOO21516.
This work was also supported by the Department of Energy Office of Science, Office of Advanced Scientific Computing Research under the guidance of Dr. Hal Finkel.

\bibliographystyle{IEEEtran}
\bibliography{main}

\clearpage

\end{document}